\documentclass[preprint2,subfigure,epsfig]{aastex}
\input{epsf}
\def\max{{\rm max}}
\def\planck{P{\sc lanck} }

\def\alm{a_{\ell m}}
\def\Ylm{Y_{\ell m}}

\def\Cl{C_{\ell}}

\begin{document}
\newcommand{\bedm}{\begin{displaymath}}
\newcommand{\eedm}{\end{displaymath}}
\newcommand{\be}{\begin{equation}}
\newcommand{\ee}{\end{equation}}
\newcommand{\etal}{{et~al.~}}
\flushbottom

\shortauthors{Naselsky}
\shorttitle{Missing parameters}

\title{MISSING PARAMETERS IN THE CMB PHYSICS}

\author{\Large{{PAVEL~D.~NASELSKY}} \\
Theoretical Astrophysics Center, Juliane Maries Vej 30,
DK-2100,  Copenhagen, Denmark.\\
Niels Bohr Institute , Blegdamsvej 17, 2100 Copenhagen, Denmark,\\
Rostov State University, Zorge 5, 344090 Rostov-Don, Russia.}


\begin{abstract}

The lecture is devoted to the comparison of a few models of cosmic recombination      
kinetics with recent CMB anisotropy data and to corresponding predictions for   
the upcoming \planck mission. The influence of additional sources of
ionized photons at the epoch of hydrogen recombination 
may restrict our possibility to estimate the density of baryonic fraction 
of matter when using \planck data. This new source of degeneracy
of the cosmological parameters which can be estimated from the CMB anisotropy 
data can mimic the distortion of the CMB power spectrum for different models of 
baryonic matter. I would like to point out that realistic values of the   
cosmological parameters can not be obtained without \planck polarization data.
\end{abstract}

\maketitle 
\onecolumn

\section{Introduction}
In my lecture, I will discuss the physics of cosmological hydrogen 
recombination and properties of the last scattering surface
(at redshift $z\simeq 10^3$), which play a crucial role for the
Cosmic Microwave Background (CMB) anisotropy and polarization formation.        
A detailed understanding of the ionization history of the cosmic plasma is      
very important in the context of recent CMB experiments, such as the BOOMERANG  
(de Bernardis et al. 2000),  MAXIMA-1 (Hanany et al. 2000), DASI (Halverson et
al.  2001), CBI (Mason et al. 2002), VSA (Watson et al. 2002) and especially, the
 new DASI polarization   
data (Kovac et al,2002 ). I will show how these experiments, including the
upcoming MAP and \planck data, would be fundamental for our understanding of     
the most general properties of the structure and evolution of the Universe,     
the history of the galaxies and the large-scale structure formation as well.    
Moreover, the \planck mission will be able to measure the
polarization of the CMB with unprecedented accuracy. I will show that
polarization power spectra contain all information about the kinetics of 
hydrogen recombination and allow us to determine the parameters of the
last scattering surface and the ionization history of the cosmic plasma at very
high redshifts ($z\sim 10^3$).

In the framework of the modern theory of the CMB anisotropy and polarization
formation the kinetics of hydrogen recombination are assumed to be ``standard''
ones. The classical theory of the hydrogen recombination
was developed by Peebles (1968), Zel'dovich, Kurt and
Sunuyaev (1968) for the pure baryonic cosmological model of the Universe and  
was generalized by Zabotin and Naselsky (1982), Jones and Wyse (1985), Seager,  
Sasselov and Scott (1999), Peebles, Seager and Hu (2000) for non-baryonic dark  
matter in the Universe. This standard model of recombination has been modified  
in various ways. 

Naselsky (1978) and Naselsky and Polnarev (1987) investigated the 
evaporation of primordial black holes as a possible source of the hydrogen
recombination delay. Avelini et al. (2000), Battye et al. (2001), Landau et al.
(2001) have pointed out that possible slow variations in time on the fundamental
constants could be important for the ionization history of the cosmic plasma.   
Sarkar and Cooper (1983), Scott et al. (1991), Ellis et al. (1992), Adams et al.
(1998) and Doroshkevich \& Naselsky (2002) have investigated the 
influence of the decays of possible unstable particles on the kinetics of
hydrogen recombination and have shown that these particles could be powerful   
sources for the distortions of the hydrogen ionization fraction. It is worth noting
 that all the models mentioned above show that the injection of non-thermal high
 energy electrons and/or photons produces the delay and distortions of the     
ionization history of the Universe starting from very high redshift $z\sim 10^3$
 and down to redshift $z\simeq 5-10$ when the galaxies started
to form. These non-standard models on the ionization history of hydrogen 
recombination can be characterized by a few additional parameters which we
needs to take into account in the framework of the best-fit cosmological
 parameter determination while using recent and future CMB anisotropy and
polarization data. I call these parameters as ``missing parameters'' of the
CMB anisotropy and polarization formation theory and I will show how important  
these parameters are for the CMB physics.  
 The aim of my talk is to compare the possible manifestation of the more        
complicated ionization history of the Universe with the contemporary CMB        
anisotropy observational data (BOOMERANG, MAXIMA-1, CBI and VSA) in order
to restrict some parameters of the models and to find the best-fit model of the 
hydrogen recombination, differing from the standard one. I'll show that a set of  
new parameters related with a more complicated ionization history of the Universe        
could play a significant role for the upcoming MAP and \planck data. 

\section{Big Bang and the general properties of the CMB physics}
Our knowledge about the Universe is based on the experimental data and 
the well-developed theory, which we call the Bing Bang model: the    
inflation paradigm, baryogenesis, the theories of cosmological
nucleosynthesis of the light chemical element, the formation of the CMB black   
body spectrum by Compton scattering of photons   
and electrons, the theories of hydrogen recombination and gravitational     
instability of the matter in an expanding Universe for the formation of
large-scale structure of the Universe. These models serve as a background of the
Big Bang theory and play an important role in the modern cosmology. Starting
with  a very hot and dense phase, the Universe expanded and cooled down. The    
black body spectra of the CMB and, in particular, the angular anisotropy and    
polarization
preserve information about early epochs of cosmological evolution. When the 
temperature of the cosmic plasma dropped down to $T\simeq 3000 K$ the free      
electrons recombined with protons into neutral hydrogen atoms and        
primordial hydrogen-helium plasma became transparent to the CMB photons. We     
call the corresponding redshift $z_r\simeq 10^3$ as the redshift of the ``last       
scattering surface'', which is the epoch $t_r$ when the Thomson optical depth   
$\tau_T \simeq 1$. The recombination ceased when the optical depth dropped down  
to $\tau_T\ll 1$ during the corresponding time interval $\Delta t_ r$ where     
$\Delta t_ r/t_r\simeq|3/2 \Delta z_r/z_r |\sim 0.1$.

The questions arise as how exactly the cosmological plasma became neutral and   
how exactly it became transparent to the CMB photons. The answers to these         
questions lie in the physics of the CMB anisotropy and polarization formation. 
  
The angular anisotropy and polarization of the CMB radiation
are related to primordial inhomogeneity of the cosmic plasma just at the
epoch of the hydrogen recombination \footnote{see the lecture notes by A.
Lasenby, J. Smoot and A. Melchiorri in this proceeding.} 
Before this epoch ($z> z_r$) the optical depth of the plasma
had been extremely high, so any ripples in the CMB temperature distribution 
caused by the primordial adiabatic perturbations of the metric, density and     
velocity  of the plasma had been practically erased. After the epoch of         
recombination the mater component in the Universe became transparent
and the CMB photons travel through the space and time from the last scattering  
surface to the observer. In Fig.1
I show the image of the CMB anisotropy from a small path of the last scattering 
surface taken by the MAXIMA-1 balloon-borne experiment.

Actually, when we look at the last scattering surface, we see the processes 
of the structure formation just at the beginning.
Small density and velocity perturbations are just traveling compressions
and decompression of the photon-electrons gas similar to sound waves. These 
cosmological sound waves produce the ripples in the CMB map, which corresponds  
to the hot (white) and cold (black) spots in Fig.1.
\begin{figure}[!t]
\centering
\vspace{0.01cm}\hspace{-0.1cm}\epsfxsize=7cm
\epsfbox{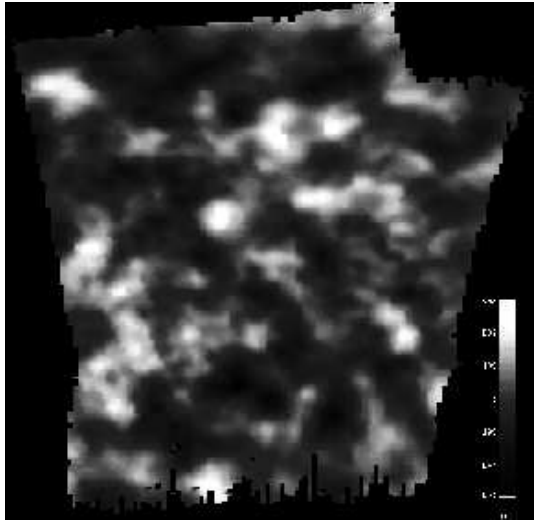}
\vspace{0.5cm}
\caption{The CMB image of the last scattering surface produced by the MAXIMA-1
experiment.}
\end{figure}
The cosmological sound waves are quite special. Before recombination
the sound speed had been close to $1.7 \times 10^4$ km/s, but right after
recombination it became 3 km/s! Such drastic transformation of the properties of the   
matter was described by Andre Sakharov in 1965 in connection with the      
matter density perturbations. The corresponding prediction by Sakharov is as     
follows. The amplitude of the perturbations after transition should preserve    
the modulation from the epoch before the transition should be preserved
Because of the very high speed of sound the typical wavelength of the 
perturbations is close to the size of the horizon at the epoch of recombination.
It corresponds to the scale of 300 Mpc at the present moment of the expansion.
The ratio between the spatial scale of the fundamental mode, which is $\sim$ 300
 Mpc, and the present horizon scale is exactly the angular measure of the CMB   
anisotropy perturbations, which is close to 30 arcmin. 

In discussion which follows, I will use a few characteristics of the CMB        
anisotropy. Firstly, let me introduce the fluctuations of the temperature
$\Delta T$ in the spherical coordinate system with polar angle
$\theta$ and azimuthal angle $\phi$.

We can decompose $\Delta T(\theta, \phi)$ as a function of $\theta$ and
$\phi$ with coefficients $\alm$ on a sphere using spherical harmonics $\Ylm$,
\begin{equation}
\Delta T(\theta,\phi)= ~\sum_\ell \sum_{m=-\ell}^{m=\ell}~\alm~\Ylm(\theta,\phi)
\end{equation}

The coefficients $\alm$ are connected with the amplitudes and the phases
of initial metric, density and velocity perturbations, which are the raw      
material for galaxy formation.
This means that the initial conditions of the galaxies formation trough the 
coefficients $\alm$ are related to the $\Delta T(\theta, \phi)$ distribution    
in the radio sky. 

Using $\alm$ we can define the power spectrum of the anisotropy and
the perturbation of the temperature $\Delta T(\ell)$, which corresponds to      
the given value of the multiple number $\ell$.
\begin{equation}
\langle \alm a^{*}_{\ell'm'}\rangle=\Cl \delta_{\ell,\ell'}\delta_{m,m'}
\label{er}
\end{equation}
As one can see from Eq~(\ref{er}), for random Gaussian processes there are no
correlations between different harmonics and the phases of the multipoles are
uniformly distributed in the range between 0 and $2\pi$. 

In Figure 2 I show the numerical calculations
of the CMB power spectrum for various cosmological models, which are reproduced
from W. Hu's web site\footnote{http://background.uchicago.edu/~whu/}.  These two
panels show the dependence of the CMB power spectrum on the baryonic density    
parameter $\Omega_b h^2$ and on the vacuum density parameter $\Omega_{\lambda}
h^2$. 

\begin{figure}[!t]
\plottwo{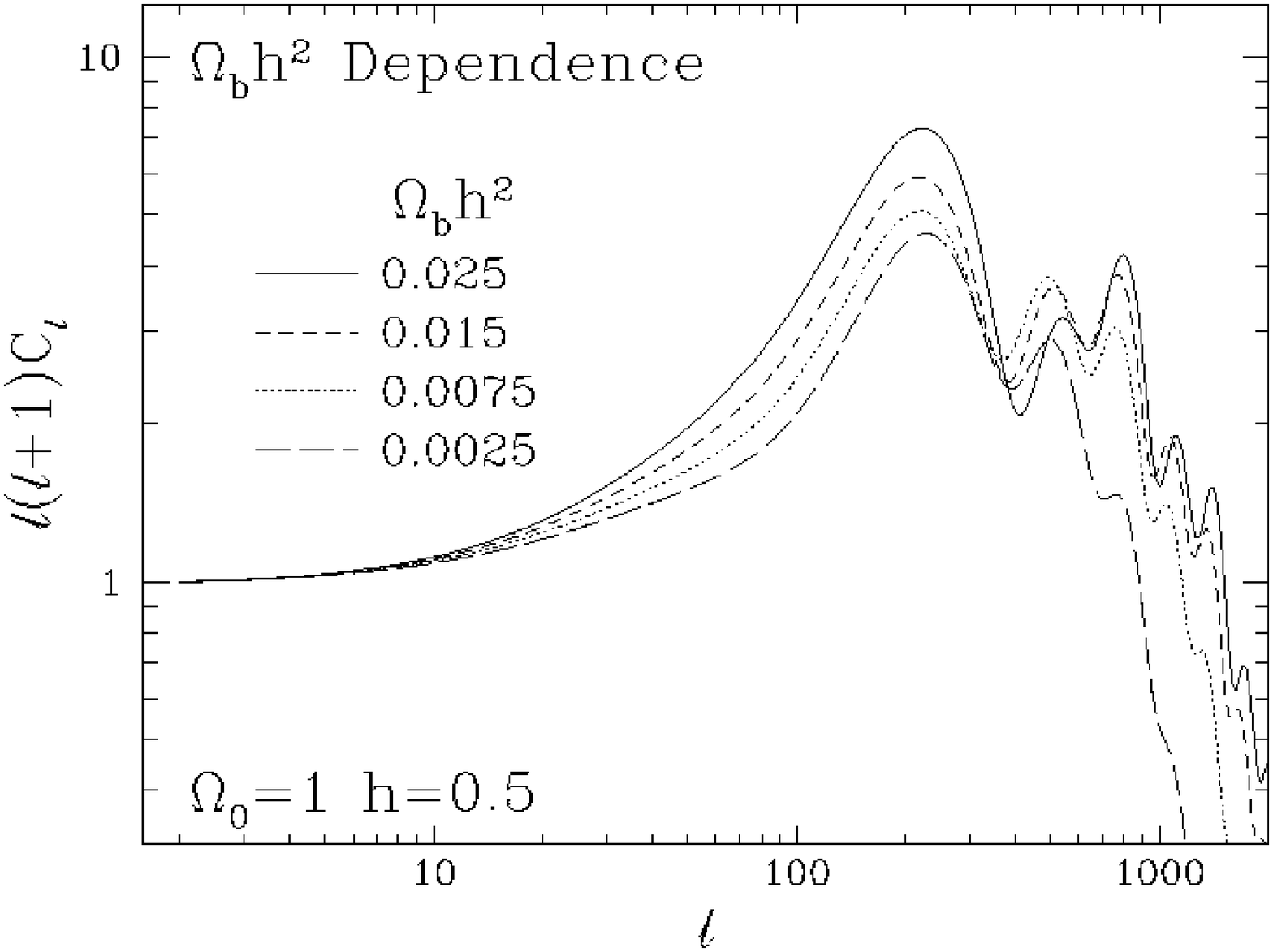}{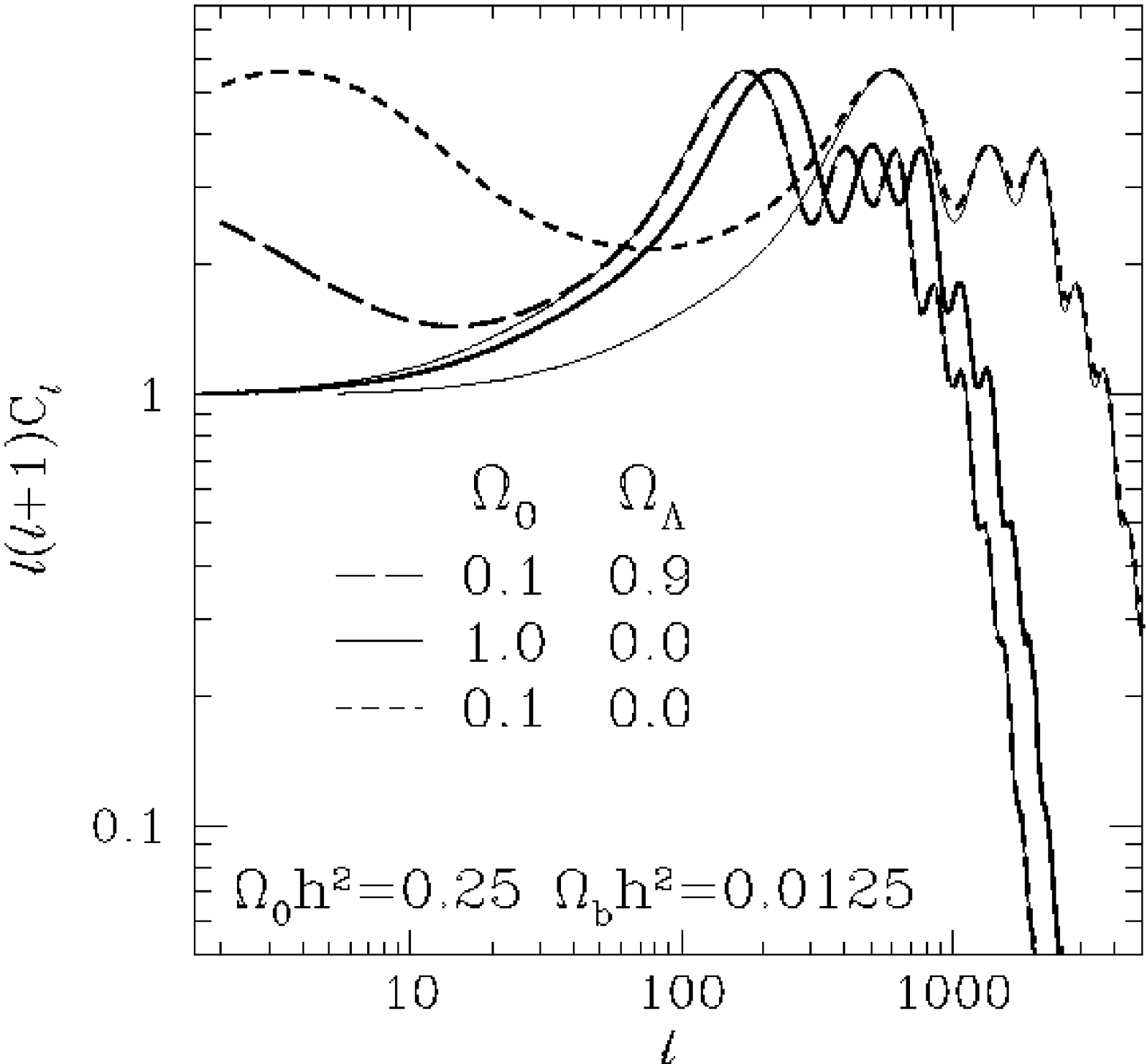}

\caption{The dependence of the CMB power spectrum upon the baryonic and vacuum
density parameter, the left and right panel, respectively. }
\end{figure}

As one can see from Fig.2 (a),  the shape of the power spectrum looks
like a set of peaks with fixed positions. For a flat universe the first peak     
corresponds to multiple number $\ell \simeq 200$.
The second one $\ell \simeq 400$, the next one $\ell \simeq 600$ and so on. 

This is the  manifestation of the sound waves in the power spectrum,  which was 
predicted by Sakharov in 1965 and was widely discussed in the literature 
related to the formation of the CMB anisotropy and polarization power spectrum. 
In Fig.2 (b) the dependence of the position and the amplitudes of the CMB          
anisotropy power spectrum peaks have shown for the $\Lambda$- and open CDM      
cosmological models  for the given values of baryonic density $\Omega_b
h^2=0.0125$. As one can see from this Figure, the CMB anisotropy power spectrum 
reflects directly all peculiarities of the $\Lambda$CDM cosmological model
related with the geometry of the Universe at 
the redshifts $z\le 0.5$. 

So, taking the above-mentioned properties into account,  we can conclude that
all peaks of the CMB anisotropy power spectrum form the basis for a new era of
the so-called  ``precision cosmology''. Namely,  major attentions of the COBE,
BOOMERANG, MAXIMA-1, DASI, VSA, CBI experiments focus on determination of
the power spectrum of the CMB  anisotropy  and testing of the statistical
properties of the signals. In addition, all experimental teams have reported
on the best-fitting cosmological models, which
result in the most probable values of the baryonic matter density $\Omega_b$, 
the density of the cold dark matter $\Omega_{cdm}$, the density of the vacuum 
$\Omega_{\Lambda}$, possible tilt of the power spectrum of the adiabatic
perturbation $n_s$, the optical depth of reionization $\tau_r$, the curvature
parameter $\Omega_K$ and so on, using most probable value of the Hubble constant
$h \equiv H_0/(100 {\rm km} \,{\rm sec}^{-1} {\rm Mpc}^{-1}) \simeq 0.65-0.70$. 
It is necessary to note that these values of cosmological parameters are
obtained from very time-consuming methods, which include as a rule 6 or 11 and
more variables such as $\Omega_b,\Omega_c, \Omega_K \ldots \tau_r, h$ and so on
(e.g. see Tegmark et al. 2001). These parameters describe the most general
properties of the Universe starting from very early epoch and down to the
present moment of the expansion. 

An intriguing question is why we need to use 6, 11 or more parameters for
determination for the most realistic cosmological model. The best explanation
for high-dimensional space of the cosmological parameters has been suggested by 
Yu. N. Parijskij. Let us suppose that we have only one peak in the CMB power
spectrum, i.e. the first peak at $\ell \sim 200$ (see Fig.2 a,b). For qualitative 
description of the peak we need three parameters: the amplitude of the peak, its
width and location. For two peaks in the power spectrum, obviously, we need to
introduce extra 3 parameters and for three peaks another 3 parameters and so on.
From Fig.2 we clearly see that at the multipole range $\ell \le 1500$  the
number of peaks in the CMB power spectrum is $N=5$ and thus corresponding number
of independent parameters should be 15. So, generally speaking, the space of the
cosmological parameters should be high-dimensional because of the peak-like
shape of the CMB anisotropy power spectrum. 

Let me remind you very briefly of some general properties of the cosmological   
parameter determination from the CMB anisotropy data sets. The parameter       
dependence leads to the following definition of the CMB power spectrum for the  
initial adiabatic perturbations,  
\begin{equation}
\Cl=C\left(\Omega_b,\Omega_{dm},\Omega_{\Lambda},\Omega_K,\Omega_{\nu},
\Omega^m_{\nu},n_s,n_t, r,\tau_r\ldots\left[i\right]\right)
\label{eqp}
\end{equation}
where $\Omega_{dm}$ is the dark matter density (including cold and warm
components) scaled to the critical density of the matter $\rho_c$, $\Omega_K=   
 1-\sum_j \Omega_j$ is the curvature parameter, which determines the flat,  
close or open model of the Universe, $\Omega_{\nu}$ and  $\Omega^m_{\nu}$
are the density of massless and massive neutrino, $n_s$ is the spectral index   
for adiabatic perturbations (e.g. $n_s=1$ for the Harrison-Zeldovich power
spectrum), $n_t$ is the corresponding spectral index for the gravitational
waves, $r=C^t(\ell=2)/C^s(\ell=2)$ is the ratio of the gravitational waves and
scalar $\Cl$ power spectrum at $\ell=2$, $\tau_r$ is the optical depth of the     
hydrogen-helium reionization caused by the first quasar and galaxy activity. 
The notation $[i]$ in Eq~(\ref{eqp}) is the main topic of my lecture, meaning that 
some  additional parameters resulting from the history of hydrogen
recombination could play a significant role in the transition of ``the CMB power spectrum
$\rightarrow$  cosmological parameters''. Obviously, for the standard
 model of recombination  we assume $[i]=0$, which means
that we do not need any additional parameters in  Eq~(\ref{eqp}). If the
kinetics of hydrogen recombination is distorted, however, by any sources of
ionization or have some peculiarities of the spatial distribution of baryonic
fraction, then it is necessary to include a few new parameters to the standard  
scheme. The question is how many parameters can characterize the more
complicated ionization history of hydrogen recombination at redshifts $z\le
1000-1500$? In order to answer this question it is necessary to examine in
details the ``standard model'' of hydrogen recombination.

\section{The standard recombination: an approximate model}
The recombination of free electrons with protons into neutral hydrogen
atoms during the cosmological expansion means that atomic physics, and more
specifically, the physics of hydrogen and helium atoms, contain some
characteristics of the recombination process, which does not depend on the
cosmology and can manifest themselves in a laboratory. Obviously, the
corresponding atomic parameters are related with the structure of the electron  
energy levels in hydrogen atoms.\footnote{I will discuss the kinetics of
hydrogen recombination as a mean process of the formation of the last scattering
surface. The kinetics of the cosmological helium recombination is included in the
standard RECFAST code (Seager, Sasselov and Scott 1999) and I send off
the reader to that paper.}
Fortunately for the physics of the CMB anisotropy formation, the physics of
the electronic transition rates in hydrogen is well known from theoretical as
well as experimental point of view. The frequency $\nu_{n, n^{'}}$ of a photon
emitted or absorbed in a transition between two levels of energy in hydrogen
with principle quantum numbers $n, n^{'}$ is given by well-known equation from quantum mechanics
\begin{equation}
E=h\nu= E_n - E_{n^{'}},
\label{e}
\end{equation}
where $h$ is the Planck constant (not the dimensionless Hubble constant!), and
$E_n=Ry \, n^{-2}$, $Ry=e^2/2a_0=13,6 $eV, $a_0={\hbar}^2/me^2$.
To liberate a free electron from a neutral hydrogen atom one needs a photon at  
least the energy $E=h\nu_{n, n^{'}\rightarrow \infty}=\chi_n$, where
$\chi_n$ is the ionization potential from the initial state $n$. For $n=1$
the corresponding potential is exactly $I=Ry=13,6 $eV and $ \chi_n$ decreases
as $n^{-2}$ for $n\ge 1$. But it is exactly what we have in the cosmology
just before the hydrogen recombination!

As mentioned in the Introduction, the ``hot'' Big Bang model of the Universe
indicates that the most widespread component in our Universe is the CMB photons, 
which has a perfect black-body spectrum and the corresponding number density of 
the photons $n_{\gamma}\simeq 420{\rm cm}^{-3}$ at the present
time of the cosmological evolution. In comparison, the number density of baryons
 $n_b\simeq 2 \times 10^{-7}(\Omega_b h^2/0.02)$ is significantly smaller than
the CMB photons. The corresponding photon-baryon ratio $s=n_{\gamma}/n_b \simeq 2
\times 10^9 (\Omega_b h^2/0.02)^{-1} $ is an excellent measure how hot the      
Universe is. Taking into account the temperature $T_0=2.728$ from the COBE data 
for the black-body CMB spectrum (Bennet et al. 1996), we can find the number
density of the $Ly_{\alpha}$ quanta ($I_{\alpha}=10.2 {\rm eV}$) in the Wien
regime of the spectrum $n(E)$ at the present time,
\begin{equation}
n_{\alpha}=\int_{I_{\alpha}}^{\infty}
 n(E)dE\sim n_{\gamma}\exp\left(-\frac{h\nu_{\alpha}}{kT_0}\right)
\label{eqw}
\end{equation}
where $I_{\alpha}=h\nu_{\alpha}$ is the energy of the $Ly_{\alpha}$ quanta 
and $k$ is the Boltzmann constant. If we reverse the cosmological expansion and
trace back in time the temperature of the black-body radiation increases  and in
 term of the redshifts it corresponds to $T(z)=T_0(1+z)$. When the redshift $z$
is close to the redshift of the last scattering surface ($z_r$), then
$T(z_r)\simeq 2700$ K and the corresponding energy $E=kT(z_r)\simeq 0.3$ eV.
Thus, the number density of the $Ly_{\alpha}$ quanta at $z=z_r$ is $
n_{\alpha}(z_r)\sim 2 \times 10^{-15}n_{\gamma}$ and increase rapidly for the   
higher redshifts and reach the number density of baryons at $z\sim 1500$. For   
the $Ly_{c}$ quanta with $ E= I=13.6 {\rm eV}$ the number density
$n_c\sim n_{\alpha}\exp[- (I-I_{\alpha})/kT(z)]\ll n_{\alpha}$ at   
 the redshifts $z\le 1500$ and decreases rapidly when $z\rightarrow z_r$. 
 So, taking properties of the electron transition in a hydrogen
atom into account, one can conclude that photo-ionization of the neutral
hydrogen atoms by the  $Ly_{\alpha}$ and $Ly_{c}$ quanta from the
CMB power spectra should by a one of the main process of the recombination
kinetics.

One additional remark comes from the analysis of the photo-ionization cross-sections 
for resonant $Ly_{\alpha}$ and $Ly_{c}$ quanta with neutral hydrogen atoms.
In the framework of the theory of the CMB anisotropy and polarization formation 
the last scattering surface is characterized by the Thomson cross-section 
$\sigma_T=8\pi/3 (e^2/mc^2)^2\simeq 6.65 \times
10^{-25}  cm^2$, where $e$ and $m$ are the charge and the mass of an electron, 
and the corresponding optical depth $\tau_T\simeq \int \sigma_T n_e c dt$. 
For the photo-ionization of 
neutral hydrogen atoms which forms during the epoch of hydrogen recombination 
the corresponding optical depth
\begin{equation}
\tau_{ph}(\nu) = \int \sigma_{ph}(\nu) n_{H_i} c dt 
 \sim  \frac{\sigma_{ph}
(\nu)}{\sigma_T}x^{-1}_{e,i}
\tau_T|_{z=z_r}\gg \tau_T|_{z=z_r}\sim 1
\label{eqt}
\end{equation}
where $n_{H_i}$ is the number densities of excited hydrogen atoms at different 
energy levels of the electrons,
$\sigma_{ph}(\nu)$ is the corresponding cross section of photo-ionization from  
the excited  energy levels of the electrons to
the continuum, $c$ is the speed of light and $x_{e,i}=n_e/n_{H_i}$. In the
quantum theory of the spectral lines we have the following normalization of the 
photo-ionization cross sections (see for the reviews by Rybicki and Lightman 1979)
\begin{equation}
\int_0^{\infty}\sigma_{ph}(\nu)d\nu= \frac{\pi e^2}{mc} f_{n\rightarrow n^{'}}
\label{eqt1}
\end{equation}
where $f_{n\rightarrow n^{'}}=f_{nn^{'}}$ is called the oscillator strength  
for the transition between states $n$ and
$n^{'}$. For example, the $Ly_{\alpha}$ transition $(n=1,n^{'}=2)$ in hydrogen 
yields the $f_{1,2}$ value  $f_{1,2}=0.4162$  (Rybicki and Lightman 1979). Thus,
we can conclude that for the  $Ly_{\alpha}$ transition the 
corresponding optical depth in the center of the line is approximately more that
10 orders of magnitude higher than the
Thomson optical depth at the redshift $z_r$. For the bound- free transition for 
hydrogen corresponding cross-section
is given by the formula ( Rybicki and Lightman,1979):
\begin{equation}
\sigma_{bf}= \left(\frac{64\pi n g}{3\sqrt3 }\right)\alpha a^2_0\left(\frac{\omega_n}
{\omega}\right)^3
\label{eqt2}
\end{equation}
where $g$ is the Gaunt factor, $\alpha=e^2/\hbar c=1/137$ is the 
fine-structure constant, $a_0={\hbar}^2/me^2$
and $h \omega_n=I_n$, $I_n={\alpha}^2 mc^2/2n^2$ is the ionization 
potential for the $n\rightarrow $ continuum
 transition. For $Ly_{c}$ quanta the corresponding cross-section is $\sigma_I\sim 
{\alpha}^{-3}\sigma_T\sim 2 \times 10^6 \sigma_T$ and the optical depth is
approximately 6 orders of magnitude higher than the Thomson 
optical depth at the moment of recombination. So, as one can see from the above-
mentioned properties of the photo-ionization cross section, the CMB photons play
 an important role in the kinetics of hydrogen recombination in the Universe.
Following Zeldovich, Kurt and Sunuyaev (1968) and Peebles (1968), let us discuss
the quantitative model of this process during the epoch $z<1500$.

\subsection{ Approximate theory of hydrogen recombination} 
As it was mentioned above in the expanding Universe the free electrons recombine
with protons into neutral hydrogen atoms when the temperature of the
matter-photon fluid drops down to $T\sim 3000-4000$ K. 
As the first step of our investigation on the recombination process we have
considered the distribution among the levels of a single atom in thermal 
equilibrium. We want to determine the distribution of an atomic species
among its various stages of ionization, which is described by the well-known Saha
 equation.  This equation describes the early stages of hydrogen recombination. 
Using the so-called three levels of the electron-energy model of hydrogen atom
as shown in Fig.3, we can start to describe the transition of the electron from 
the ground level to the continuum in thermal equilibrium (Rybicki and
Lightman 1979). 

\begin{figure}[!t]
\centering
\vspace{0.01cm}\hspace{0.1cm}\epsfxsize=5cm
\epsfbox{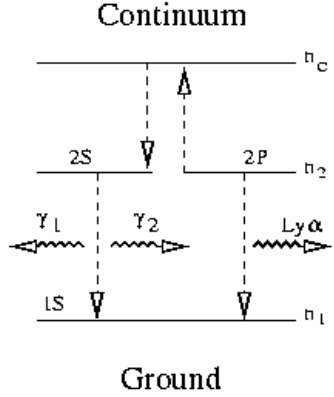}
\vspace{0.5cm}
\caption{The three electron energy levels model of the hydrogen atom
. }
\end{figure}
According to the Boltzmann law the differential number of ions at the ground level 
$(N^+(v))$ with the free electrons at the velocity range between 
$v$ and  $v+dv$ is
\begin{equation}
\frac{N^{+}(v)}{N_0}=\frac{g}{g_0} \exp\left[-\frac{I+\frac{1}{2}mv^2}{kT}\right] 
\label{eqs}
\end{equation}
where $N_0$ and $g_0$ are the number and statistical weight of the hydrogen
atoms at the ground state, $I=13.6 eV$ is the ionization potential, 
$g=g^{+}_0g_e$ is the product of the statistical weight of the ion at the ground
state $(g^{+}_0)$ and the differential electron statistical weight
$g_e=2d^3\vec{x}d^3\vec{p}/h^3$. (Note that the factor 2 in $g_e$
reflects the two spin states of the electron.) 
The volume element $ d^3\vec{x}=n^{-1}_e$, where $n_e$ is the number density of  the 
free electrons, and the volume element $d^3\vec{p}$ is equivalent to $4\pi m^3
v^2dv$ assuming isotropic velocity distribution. 
Using Eq.~(\ref{eqs}) after integration over all $v$ we obtain 
\begin{equation}
\frac{n_p n_e}{n_0}=2\frac{g^{+}_0}{g_0}\left(\frac{2\pi m kT}{h^2}\right)^3\exp 
(-I/kT)
\label{eqs1}
\end{equation}
where $n_p$ is the number density of protons and $n_0$ is the number density of 
the hydrogen atoms at the ground level.
As one can see from  Eq.~(\ref{eqs1}) the fraction of free electrons decreases  
very fast during the cosmological expansion because
of the exponential factor $(I/kT)\gg 1$. However, I would like to put forward  
two important remarks while applying the Saha equation
for the kinetics of cosmological hydrogen recombination. Firstly, it is
necessary to note that during recombination the thermal equilibrium between
black-body radiation and atoms has been affected by the deviation of the
radiation spectrum from the Planckian form. Since recombination occurs for
$kT\ll I$ the number density of photons with $h\nu=I$ would be far
smaller than the number density of electrons and protons. The range of the
photon spectrum that is crucial for the recombination process actually 
changes substantially if recombining electrons proceed directly to the ground
state. Secondly, it is necessary to take into account
the cascading recombination process, which plays the most important role in the 
recombination kinetics and produce corresponding delay of hydrogen recombination
 relative to the ``standard'' fast Saha recombination kinetics. The importance
of the cascading recombination process was pointed out by Zeldovich, Kurt and
Sunuyaev (1968) and Peebles (1968). 
For the cascading recombination approximation we can neglect the direct
recombination to the ground state but introduce the net recombination to the 
level ($n,l$),
\begin{equation}
-\frac{d}{dt}\left(\frac{n_e}{n_b}\right)=n^{-1}\sum_{n>1}\left(\alpha_{nl}
n^2_e -\beta_{nl}n_{nl}\right)
\label{eqe}
\end{equation}
where $\alpha_{nl}$ is the recombination coefficient to the ($n,l$) level of energy, 
$n_{nl}$ is the
number density of the excited hydrogen atoms,  $\beta_{nl}$ is the rate of
 photo-ionization of a hydrogen atom at this level and $n_b$ is the number
density  of baryons at redshift $z$. Assuming that the process of recombination 
does not perturb significantly the black-body spectrum of the CMB  at $h\nu\ll I$
range, the ratio of $\beta_{nl}$ to $\alpha_{nl}$ is given by the Saha equation 
for $n\ge 2$,
\begin{equation}
\beta_c  = \sum_{n>1}(2l+1)\beta_{nl}\exp\left(-\frac{I_{\alpha}-I_n}{kT}\right) \nonumber\\
 = \alpha_c\frac{\left(2\pi m kT\right)^{3/2}}{h^3}
\exp\left(-\frac{I_{\alpha}}{kT}\right),
\label{eqe1}
\end{equation}
where $\alpha_c =\sum_{n>1}\alpha_{nl}$ and $I_n$ is the potential of ionization
 from the level $n$. In addition, it is necessary
to take into account that in equilibrium  the number density of atoms at the
level  $n,l$  satisfies the following
\begin{equation} 
n_{nl}=n_{2s}(2l+1)\exp\left(-\frac{I_{\alpha}-I_n}{kT}\right).
\label{eqe2}
\end{equation}
So, from Eq~(\ref{eqe})-(\ref{eqe2}) we obtain the well-known relation for the  
number density of the free electrons (Peebles 1968),
\begin{equation}
-n_b\frac{d}{dt}\left(\frac{n_e}{n_b}\right)=\alpha_c n^2_e - 
\beta_c G_{\alpha}n_{1s},
\label{eqe3}
\end{equation}
where $G_{\alpha}=n_{2s}/n_{1s}$ is the number of photons per mode in the 
$Ly_{\alpha}$ resonance line.

Let me describe briefly the evolution of the resonance $Ly_{\alpha}$ line at
the epoch of hydrogen recombination (Zeldovich, Kurt and Sunuyaev 1968, Peebles 
1968). One of the most important results from the $Ly_{\alpha}$ photo-ionization
 cross section is that the mean free time of the resonance recombination
line of photons becomes very short, at the order $\sim 30$ seconds while the age
 of the Universe at the redshift $z\sim 1500$
is at the order of $10^{12}$ s. That means that the optical depth for the  $Ly_{\alpha}$ 
photons at the center of the line is extremely high, at the order of $10^{10}$  
and the radiation in the line is strongly coupled with 
the formed hydrogen atoms. On the other hand, the cosmological expansion of the 
Universe redshifted the $Ly_{\alpha}$ photons from
the center of the line and decrease rapidly the cross section of photo-ionization. 
To obtain the magnitude of the redshift effect, we can use the approximate
equation for the spectrum of the electromagnetic radiation (Peebles 1968),
\begin{equation} 
n_b\frac{d}{dt}\left(\frac{n_{\alpha}}{n_b}\right)=\nu_{\alpha}H(t)(n_{\nu_{\max}}-
n_{\nu_{\min}}) + R(t),
\label{eqla}
\end{equation}
where 
\begin{equation}
n_{\alpha}=\int_{\nu_{\min}}^{\nu_{\max}}n(\nu)d\nu
\label{eqla1}
\end{equation}
and $H(t)$ is the Hubble parameter,  $\nu_{\min}$ and $\nu_{\max}$ are the minimal
 and maximal frequencies just below and above the center of the line,
$n(v)$ is the spectral number density of the photons in the line, $R(t)$ is the net
 rate of production of the $Ly_{\alpha}$ photons
per unit volume during $2p\rightarrow 1s$ transition. In order of magnitude, taking 
 Eq.~(\ref{eqla1}) into account, we have
$n_{\alpha}\sim n_{\nu_{\min}}\Delta\nu $, where $\Delta\nu $  
is the  $Ly_{\alpha}$ line width and $\Delta\nu/\nu_{\alpha}\sim 10^{-5}$. Thus,
 from  Eq.~(\ref{eqla}) one can obtain (Peebles 1968)
\begin{equation} 
n_{\nu_{\min}}=n_{\nu_{\max}} + R\nu_{\alpha}H^{-1}(t)
\label{eqla3}
\end{equation}
Using dimensionless variable $ G=c^3 n(\nu)/8\pi {\nu}^2$, we can change the last
equation to the following form
\begin{equation}
G_{\alpha}= G_{+} + R {\lambda^3_{\alpha}}H^{-1}(t)/8\pi,
\label{eqla11}
\end{equation}
where $G_{+}$ corresponds to $n(\nu_{\max})$. 
Because of the intimate contact with the atoms the number of photons per mode 
satisfies the equation  
$G_{+}=\exp[-(I_{\alpha}-I_n)/kT]$ and for $G_{\alpha}$ in 
Eq.~(\ref{eqla1}) we get
\begin{equation}
G_{\alpha}= \exp\left(-\frac{I_{\alpha}-I_n}{kT}\right) +
R\nu_{\alpha}H^{-1}(t).
\label{eqla4}
\end{equation}
The last point of the approximate model of hydrogen recombination is related 
to the estimation of the net recombination $R(t)$, which defines the ionization 
fraction of the hydrogen. Thus
\begin{equation} 
R(t)= \left[\alpha_c n^2_e - \beta_c G_{\alpha}n_{1s}\right] - 
\Lambda_{2s,1s}\left[n_{2s}-n_{1s}\exp\left(-\frac{I-I_{\alpha}}{kT}\right)\right]
\label{eqla5}
\end{equation}
where $\Lambda_{2s,1s}=8.227 {\rm sec}^{-1}$ is the decay rate from $2s$ state.
The excess of recombination over photo ionization results in the production of  
the $Ly_{\alpha}$ quanta (the first term in Eq.~(\ref{eqla5})) or in a
two-quantum decay of the meta-stable $2s$ level of energy of the hydrogen atom  
(the last term in Eq.~(\ref{eqla5})). So, after simple algebra from
Eq.~(\ref{eqla})-(\ref{eqla5}) we obtain the well-known equation for the free    
electron number density (Peebles 1968):
\begin{equation}
 -n_b\frac{d}{dt}\left(\frac{n_e}{n_b}\right)=C\left[\alpha_c n^2_e - 
\beta_c n_{1s}\exp\left(-\frac{I-I_{\alpha}}{kT}\right)\right],
\label{eqla6}
\end{equation}
where 
\begin{equation}
C=\frac{1+K\Lambda_{2s,1s}n_{1s}}{1+K(\Lambda_{2s,1s}+\beta_c)n_{1s}},
\end{equation}
and $K={\lambda^3_{\alpha}}H^{-1}(t)/8\pi$. 
It is necessary to note that the approximate model of hydrogen recombination
 assumes that the temperature of the plasma $T_m$ during the recombination
process is equivalent to black-body radiation temperature and the contribution of   
the helium atoms is negligible. A more general and detailed
model of recombination was developed by Seager, Sasselov and Scott (1999), which
includes the Compton interactions between electrons and photons and the
corresponding deviation $T_m$ and $T_{cmb}$, recombination of helium and
ionization by the electrons. The qualitative theory of the hydrogen 
and helium recombination was summarize in the RECFAST program, which was
incorporate in the CMBFAST code (Seljak and Zaldarriaga 1996) for calculations  
of the power spectrum of the CMB anisotropy and polarization for various
cosmological models. Below we will use the modification of the the RECFAST and  
CMBFAST codes to investigate more complicated than ``standard'' recombination regimes.

\section{Delayed and accelerated recombination}
As it was mentioned in Introduction and Section 3, the standard model of   
hydrogen recombination assumes that at the epoch $z\le 1500$ the production of  
the resonant $Ly_{\alpha}$ and $Ly_{c}$ photons is determined by the black-body 
spectra of the CMB and transitions 
of electrons into the neutral hydrogen atoms. Generally speaking, any decay of  
some relics of the cosmological evolution
of matter in the Universe at the epoch of recombination are negligible. This
assumption, however,  needs to be
proven by recent and future CMB experiments, especially by the  \planck observation.

 As it was mentioned in Introduction a lot of non-standard models   
have been developed during last few years especially for the upcoming \planck experiment, 
which is characterized by unprecedent accuracy of the anisotropy and
polarization power spectrum determination . 
The aim of this section is to compare the possible manifestation of the more 
complicated ionization history of the Universe with the contemporary CMB anisotropy
observational data  (BOOMERANG, MAXIMA-1, CBI and VSA) in order
to put constraints on some parameters of the models and to classify the  models
of hydrogen recombination, differing from the standard one. 
   
The basis for classification of any models of delayed and accelerated recombination
is related with  the model proposed by Peebles, Seager and Hu (2000), in which
two independent parameters $\varepsilon_{\alpha}$ and $\varepsilon_{i}$ are
introduced as effectives of the non-thermal resonant $Ly_{\alpha}$ and ionized-photon production
before, during and after the period $z\simeq 10^3$,  
\begin{equation}
\frac{dn_{\alpha}}{dt}=\varepsilon_{\alpha}H(t)n_H,
\end{equation}
 and 
\begin{equation}
\frac{dn_{i}}{dt}=\varepsilon_{i}H(t)n_H,
\end{equation}\label{eq1}
where $n_H$ is the fraction of neutral hydrogen.
In their model the parameters  $\varepsilon_{\alpha}$ 
and $\varepsilon_{i}$ are assumed to be constant in time, which means that
corresponding number density of the  $Ly_{\alpha}$ and ionized-photon production
 per interval $H^{-1}$ are $\Delta n_{\alpha}= H^{-1}
dn_{\alpha}/dt=\varepsilon_{\alpha}n_H$  and 
$\Delta n_{i}= H^{-1} dn_{i}/dt=\varepsilon_{i}n_H$. As one can see, if 
$\varepsilon_{\alpha}\ll 1 $ and $\varepsilon_{i}\ll 1$ then the corresponding  
distortions of the recombination kinetics should be small. It is natural to
assume that the corresponding number density of free electrons produced by
photo-ionization of the neutral hydrogen atoms is proportional to
$\Delta n_{i}$ and $\Delta n_{\alpha}$ and corresponds to the electron
ionization  fraction
$x_e\sim \Delta n_{i}/n_H\sim \varepsilon_{i}$ or 
$x_e\sim \Delta n_{\alpha}/n_H\sim\varepsilon_{\alpha}$ for different stages of 
hydrogen recombination. However, the concrete values and possible variation of  
the  $\varepsilon_{\alpha}$ and $\varepsilon_{i}$ parameters in time depends on 
the energy injection spectra of the sources of the non-thermal resonant
$Ly_{\alpha}$, ionized
photons production and on the model of the source decay. 
 
Let us suppose that all resonant and ionized
photons are the result of some massive unstable particles decay (Doroshkevich and
Naselsky 2002), which include primordial black hole evaporation. For such      
particles the number  density decreases in time as
\begin{equation}
\frac{dn_{X}}{dt}+3H(t)n_X=-\frac{n_X}{\tau_X(n_X,t)},
\label{eqd2}
\end{equation}
where $n_X$ is the number density of the $X$-particles, and $\tau_X(n_X,t)$
is the life time, which can be constant for the standard decay or as
a function of cosmological time, if the
production of the ionized and resonant quanta is related with annihilation
of the $x$ particles. After normalization $n_X=\tilde {n}_X 
\left(\frac{a_0}{a}\right)^3$,  where $a_0$ is the scale factor of the
Universe at some moment $t_0$, we can transform Eq.~(\ref{eqd2}) to the
following form  
\begin{equation}
\frac{d\tilde {n}_{x}}{dt}=\varepsilon_{x}(t)H(t) n_{H},
\label{eqq3}
\end{equation}
where 
\begin{equation}
\varepsilon_{X}= - \left(H\tau_X\right)^{-1} \exp\left(-\frac{t-t_0}{\tau_X}\right)
\left(\frac{\tilde {n}_{X,in} }{{n}_{H}}\right)
\label{eqq4}
\end{equation}
 and $\tilde {n}_{X,in}$ is the initial number density of the $X$-particles
at the moment $t_0$.

I would like to point out that Eq.~(\ref{eqq3}) has a general character.
Obviously, we can include to the function $\varepsilon_{X}(t)$ all models of the
 particle decay and then $\varepsilon_{\alpha} $ and  $\varepsilon_{i} $ should
be  proportional to $-\varepsilon_{X}(t)$. As the result, they are functions
of time (or redshift $z$).

For illustration of the tight dependency  $\varepsilon_{\alpha} $, 
$\varepsilon_{i} $ and
$\varepsilon_{X}(t)$ parameter, let me describe the model of the
Super Heavy Dark Matter (SHDM) particle decay during the cosmological expansion.
Such a model was discuss in connection with the origin of the 
Ultra High Energy Cosmic Rays (UHECRs) in our Galaxy in the framework of
so-called Top-Down scenario. As was suggested by Berezinsky, Karshelrie$\ss$ and
 Vilenkin (1997), Kuzmin and Rubakov (1998), Birkel and Sarkar (1998),
 the formation of such UHECRs can be related to decays of the various kinds of
SHDM  particles with masses $M_X\ge 10^{12}$GeV.  
This model was widely discussed in connection with AGASA (Hayashida et al. 1994;
Ave et al. 2000), Fly's Eye (Yoshida and Dai 1998; Abu-Zayyad et al. 2002) 
\footnote{see for details E. Loh's lecture in this proceeding}, Havera Park
(Lawrence, Reid  and Watson 1991) experiments.  

\subsection{Expected flux of high energy photons} 
As is commonly believed, decays of SHDM particles into the 
high energy protons, photons, electron-positron pairs and neutrinos 
occurs through the production of quark-antiquark pairs ($x\rightarrow
q,\overline{q}$), which rapidly hadronize, generate two jets and transform 
the energy into hadrons ($\omega_h\sim$5\%) and pions ($1-\omega_h
\sim$95\%) (Blasi 1999). It can be expected that later on the energy 
is transformed mainly to high energy photons and neutrinos. This means 
that, for such decays of SHDM particles with $10^{12}$GeV$<{M_x}
<10^{19}$ GeV, the UHECRs with energy $E>10^{20}$ eV are dominated 
by photons and neutrinos (Blasi 1999). This conclusion can be tested 
with further observations of the UHECR fluxes at $E>10^{20}$eV. 

\begin{figure}[!t]
\plottwo{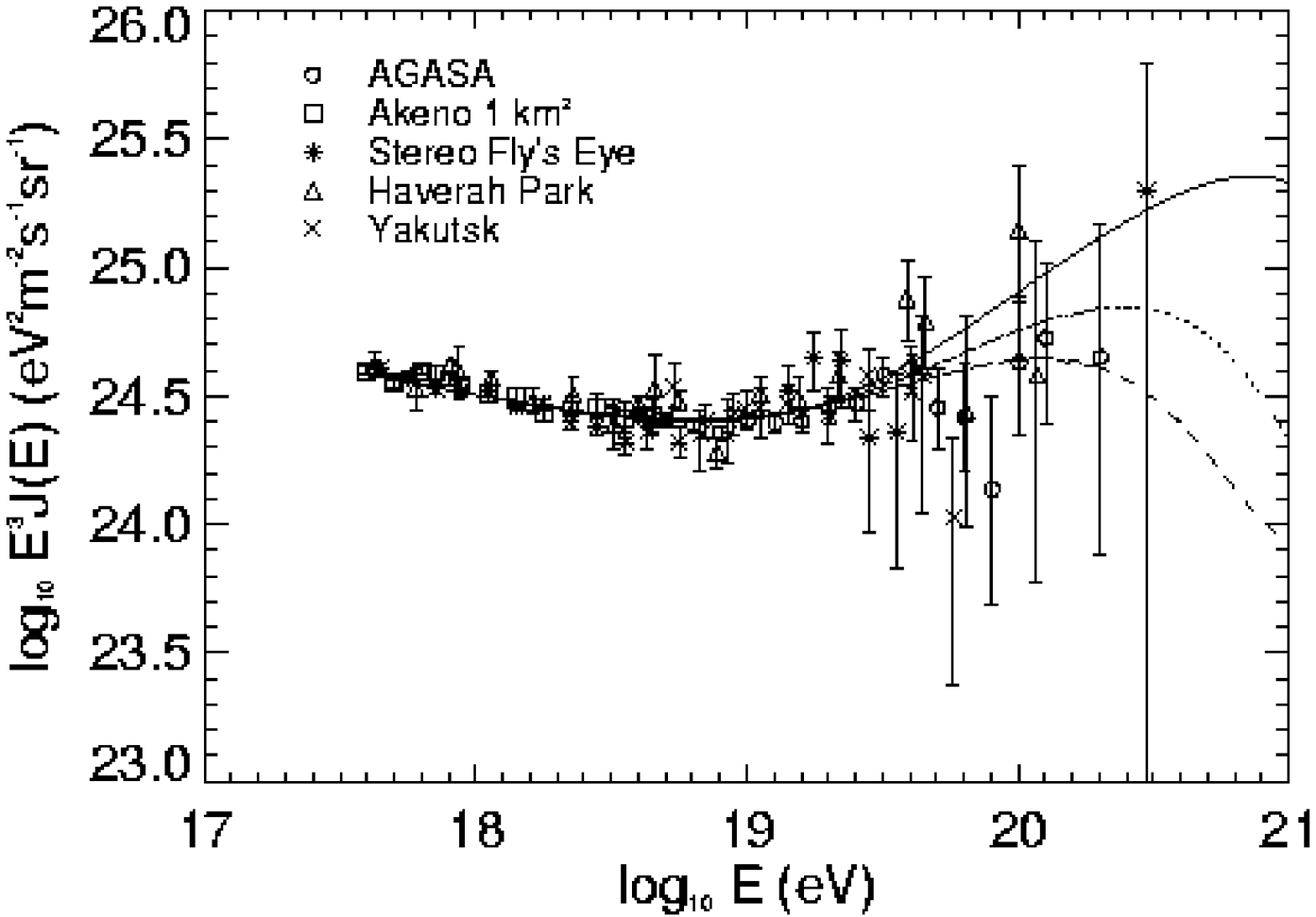}{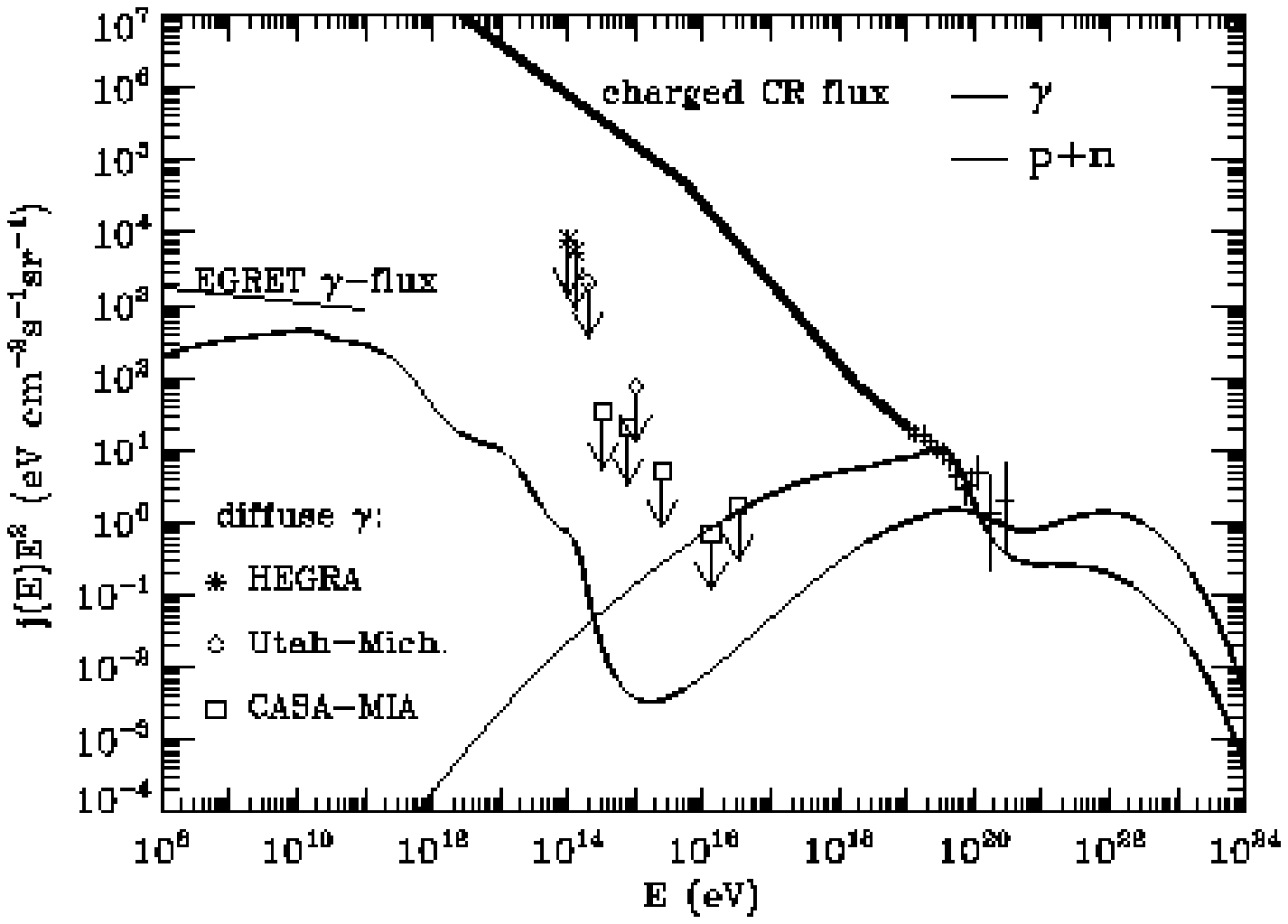}
\caption{The left panel: the observed  spectrum for the UHECR and Top-Down scenario predictions
for different $M_X$. The dash line is for $M_X=10^{11} $GeV, the dot line 
$M_X=10^{2}  $GeV, and the  solid line $M_X=10^{13} $GeV.
(From Sarkar and Toldra 2002. The right panel: the $\gamma$-rays spectra from UHECR in
comparison  with EGRET data. See for details in Sigl (2001). }
\end{figure}

Probable energy losses of neutrinos are small (Bhattacharjee and Sigl 2000)
, they are concentrated at the range $E\sim E_{inj}\gg E_{\rm GZK}$ and could be 
responsible only for relatively rare observed events. But the 
comparison of expected and observed fluxes of high energy photons 
restricts some properties of SHDM particles and the Top--Down model 
as a whole ( Berezinsky, Blasi and Vilenkin 1998). At high redshifts the
interaction  of high 
energy photons with $E\sim E_{inj}\gg E_{\rm GZK}$ with the CMB background 
($\gamma_{\rm UHECR}+\gamma_{\rm CMB}
\rightarrow e^{+} + e^{-}$)leads to formation of electromagnetic cascades
(Greisen 1966; Zatsepin  and Kuzmin 1966).

 At small redshifts 
the efficiency of this interaction decreases and the evolution of 
such photons depends upon unknown factors such as the extragalactic 
magnetic field and properties of radio background.  

The expected domination of products of decay of the x-particle by 
the high energy pions and photons follows from quite general arguments.
The cumulative contribution of the photons is bigger then the cumulative 
contribution of the hadrons by a factor of $\sim\omega_h^{-1}\gg$ 1. 
This means that, at least for the galactic component, the cumulative 
observed flux of UHECRs at energy $E\geq E_{\rm GZK}$ is also dominated 
by galactic component of photons with $E\sim E_{inj}\gg E_{\rm GZK}$. 
The discrimination of high energy photon component of UHECRs can be 
considered as the crucial test for the Top--Down models. 

Summarizing available information about the photon component of UHECRs, Bhattacharjee
and Sigl (2000) estimate the spectrum of injected 
photons for a decay of a single $X$-particle as follows:
\begin{equation}
{dN_{inj}\over dE_{\gamma}}=\frac{0.6(2-\alpha)}{M_X}
\frac{f_{\pi}}{0.9}\left(\frac{2E_{\gamma}}
{M_X}\right)^{-\alpha},~ E_\gamma\leq {M_X\over 2}
\label{eq1}
\end {equation}
where $f_{\pi}$ is the fraction of the total energy of the jet 
carried by pions (total pion fraction in terms of of number of 
particles), $0<\alpha<2$ is the power index of the injected spectrum, 
$M_X$ is the mass of the SHDM particles. For the photon path length 
$l_{\gamma}\sim 1-10$Mpc at $E_{\gamma}\sim E_{\rm GZK}$ in respect to the  
electron-positron pair creation on the extragalactic radio background
(see Bhattacharjee and  Sigl (2000); Protheroe and  Biermann 1996)
 we get for the photon flux $j_{inj}(E_{\gamma})$ 
at the observed energy $E_{\gamma}$:
\begin{equation}
j_{inj}(E_\gamma)\simeq \frac{1}{4\pi}l_{\gamma}(E_{\gamma})
\dot{n}_{X}{dN_{inj}\over dE_{\gamma}}
\label{eq2}
\end{equation}
where $\dot{n}_{X}$ is the decay rate of the $X$  particles.
For the future calculation we will use the normalization of 
$j_{inj}(E_{\gamma})$ on the observable UHECR flux which corresponds 
to normalization on the decay rate $\dot{n}_{X}$ at present time, 
$t=t_u$ (Bhattacharjee and  Sigl 2000):
\begin{equation}
\dot{n}_{X,0}\simeq 10^{-46}{\rm cm^{-3}s^{-1}}
M_{16}^{1-\alpha}\Theta_X, ~\nonumber\\
\label{eqq}
\end{equation}
where

\begin{equation}
\Theta_X\approx\frac{0.5}{2-\alpha}(\frac{0.9}{f_{\pi}})(\frac{10 {\rm Mpc}}{l_\gamma (E_{obs})})
(2E_{16})^{\alpha-3/2}\times    ~\nonumber\\         
\left(\frac {E^2_{obs} j_{obs}(E_{obs})}
{\rm 1 eV cm^{-2} s^{-1}sr^{-1}}\right)
\end{equation}
and  $M_{16}=M_X/10^{16}$GeV, $E_{16}=E_{obs}/10^{16}$GeV, 
$E_{obs}$ and $j_{obs}(E_{obs})$ are the observable range of 
the UHECR energy and flux, respectively. Note that normalization 
(\ref{eqq}) does not depend on the nature of the $x$-particles.

An additional possibility for non-equilibrium high energy photon injection to
primordial hydrogen-helium plasma at the epoch of recombination
comes from the theory of primordial black
hole evaporation (Naselsky 1978; Ivanov, Naselsky and Novikov 1994; Kotok and
Naselsky 1998; Doroshkevich and Naselsky 2002) or relatively low massive
particle decay $M_X \ll 10^{12}$ GeV (Scott, Rees and Sciama 1991; Adams,
Sarkar  and Sciama 1998). 
For primordial black holes with masses $M_{\rm pbh}\sim 10^{13}$ gram the 
corresponding life time is $\tau_{\rm pbh}\sim t_r$ and they evaporate just
at the epoch of hydrogen recombination by emitting high energy particles
with black body spectra with effective temperature $T_{\rm pbh} \sim 10^{13}-10^{14}
$K and  energy $E_{\rm pbh}\sim kT_{\rm pbh}\sim 1-10 $ GeV (Hawking 1975). As one can
see, this range of energy injection
is sufficiently smaller than for the SHDM particles with $M_X \ge 10^{12}$ GeV. 
Note that for the models from Scott, Rees and Sciama (1991) and Adams, Sarkar
and Sciama (1998) the characteristic energy of injected photons 
from neutrino decay is of the order $10-100$ eV, which is close to the
$Ly_{c}-Ly_{\alpha}$ energy range.
Taking the above-mentioned properties of the non-equilibrium photon injection
into account it is natural
to expect that practically all range of energy above  $Ly_{c}-Ly_{\alpha}$
threshold should  be very important
for investigation of the non-standard recombination kinetics.

\subsection{Electromagnetic cascades at the epoch of 
the hydrogen recombination.}
The decays of $X$-particles cannot significantly distort the 
thermodynamic of the universe at high redshifts $z\geq 10^3$ 
but this injection of energy changes the kinetic of 
recombination at $z\sim 10^3$ that leads to observable 
distortions of the power spectra of CMB anisotropy and 
polarization. To evaluate these distortions we need firstly 
to follow the transformation of high energy injected particles 
to UV photons influenced directly the recombination process. 

The electromagnetic cascades are initiated by the ultra high energy 
jets and composed by photons, protons, electron- positrons and neutrino. 
At high redshifts, the cascades develops very rapidly via interaction 
with the CMB photons and pair creation ($\gamma_{\rm UHECR}+\gamma_{\rm CMB}
\rightarrow e^{+} + e^{-}$), proton-photon pair production ($p_{\rm UHECR} 
+ \gamma_{\rm CMB}\rightarrow p^{'} + \gamma{'}+e^{+} + e^{-}$), inverse 
Compton scattering ($e^{-}_{\rm UHECR}+\gamma_{\rm CMB}\rightarrow e^{'} + 
\gamma^{'}$), pair creation ($e^{-}_{\rm UHECR}+\gamma_{\rm CMB}\rightarrow 
e^{'}+ e^{-} + e^{+} + \gamma^{'}$), and, for neutrino, electron-
positron pair creation through the Z-resonance. As was shown by 
Berezinsky et al. (1990) and Protheroe et al. (1995), these 
processes result in the universal normalized spectrum of a cascade 
with a primary energy $E_\gamma$ which can be written as follows:
\be 
N_\gamma(E, E_{\gamma})=F(E, E_{\gamma})
\left\{
\begin{array}{cc}
\sqrt{E\over E_a}&E\leq E_a\cr
1&E_a\leq E\leq E_c\cr
0&E_c\leq E\cr
\end{array} 
\right.      
\label{eq1e}
\ee
\bedm
F(E, E_{\gamma})=\frac{E_\gamma E^{-2}}{2+\ln(E_c/E_a)},\hspace{0.5cm}
\int_0^{E_\gamma}EN_\gamma dE=E_\gamma
\eedm
where $E_c\simeq 4.6\cdot 10^4(1+z)^{-1}$GeV, $E_a=1.8\cdot 10^3
(1+z)^{-1}$GeV. At the period of recombination $z\sim 10^3$ and 
for lesser redshifts both energies, $E_a$ and $E_c$ are larger 
then the limit of the electron-positron pair production $E_{e^{+},
e^{-}}=2 m_e = 1$ MeV and the spectrum (\ref{eq1e}) describes both 
the energy distribution at $E\geq E_{e^+,e^-}$ and the injection of 
UV photons with $E\ll E_{e^+,e^-}$. However, the spectrum of these 
UV photons is distorted due to the interaction of photons with the 
hydrogen - helium plasma.  

In the range of less energy of photons, $E\leq 2m_e$, and at higher 
redshifts, $z\geq 10^4$, when equilibrium concentrations of $H_{\rm I}$, 
$He_{\rm I}$ and $He_{\rm II}$ are small and their influence is negligible, the 
evolution of the spectrum of ultraviolet photons, $N_{uv}(E,z)$, 
occurs due to the injection of new UV photons and their redshift 
and Compton scattering. It is described by the transport equation 
(Berezinsky et al. 1990)
\begin{eqnarray}
{\partial N_{uv}\over \partial z}-{3N_{uv}\over 1+z}+{\partial
\over \partial E}\left(N_{uv} {dE\over dz}\right) + {Q(E,z)\over(1+z)H} = 0, ~\nonumber\\
{1\over E}{dE\over dz}={1\over 1+z}+ {c\sigma_Tn_e\over (1+z)H(z)}
\left(\frac{E}{m_ec^2}\right) 
={1+\beta_\gamma(E,z)\over 1+z}  ~\nonumber\\
Q(E,t)=\dot{n}_X N_\gamma(E,M_X)
\label{eqq1}
\end{eqnarray}
where the Hubble parameter is
\be
H(z)=H_0\sqrt{\Omega_m(1+z)^3+1-\Omega_m},
\ee
and
 $n_e\propto (1+z)^3$ 
is the number density of electrons, so, $\beta_\gamma\propto 
(1+z)^{3/2}E$. Here $Q(E,t)$ is considered as an external source 
of UV radiation. 
 
The general solution of equation (\ref{eqq1}) is (Doroshkevich and Naselsky 2002): 
\be
N_{uv}(z) =\int_z^{z_{mx}}{Q(x)\over H(x)}
{E^2(x)\over E^2(z)}\left({1+z\over 1+x}\right)^4{dx\over 1+x}, 
\label{eqw1}
\ee
\bedm
E(x)=\frac{1+x}{1+z}E(z)\left(1-{2\over 5}\beta_\gamma(E,z)
\left[\left({1+x\over1+z}\right)^{5/2}-1\right]\right)^{-1}
\eedm
where the maximal redshift, $z_{mx}$, in (\ref{eqw1}) is defined 
by the condition $E(x)=2m_ec^2$.

As is seen from (\ref{eqw1}), the Compton scattering dominates for 
$E\gg 30$keV, when 
\be
\beta_\gamma(E,z) = 44{E\over m_ec^2}\sqrt{0.3\over\Omega_m}
{h\Omega_b\over 0.02}\left({1+z\over 10^3}\right)^{3/2}\gg 1,
\label{beta}
\ee
\be
N_{uv}(z)\propto \frac{\dot{n}_X(z)N_\gamma(E,M_X)}{H(z)
\beta_\gamma(E,z)}\propto {\sqrt{1+z}\Theta_\tau(z)\over E^{5/2}} ,
\label{eqw2}
\ee

For the most interesting energy range, $E\ll 30$keV,  
$\beta_\gamma(E,z)\ll 1$, we get again
\be
N_{uv}(E(z),z)\approx {2\over 3}\frac{\dot{n}_X(z)}{H(z)}
N_\gamma(E,E_{\gamma}=M_X/2).
\label{eqw2}
\ee

It is necessary to note that the UV photons energy  spectra Eq.~(\ref{eqw2}) reflect
directly the simple idea that practically all energy of the SHDM relics transforms 
to the energy of the $\gamma$ quanta at the range $E\le 100$ GeV (at $z=0$)(see Fig4.b)
 and
than the soft part of the spectra forms due to electromagnetic cascades.  At higher
redshift, especially at $z \sim 10^3$ the electromagnetic cascades from the $\gamma$-range
of the spectra are very effective for the UV photon production because of the high
energy density of the CMB (12 orders of magnitude higher than the present energy density) 
and because of increasing of the mean energy of the CMB photons , which is in order magnitude
equivalent to 1 eV. 

\subsection{ Delay of the hydrogen recombination due to SHDM decay}

Let me briefly discuss the relationship between initial spectra of the SHDM decay and
corresponding parameters $\varepsilon_{\alpha}$ 
and $\varepsilon_{i}$ of the hydrogen excitation and ionization . As it was mentioned 
in the previous subsection, the spectra of UV photons can be normalized to the
energy density of injection $\dot{Q}(t) $ from the SHDM decay as 
\begin{equation}
N(E)\simeq \eta\frac{\dot{Q}}{E^2_a[2+\ln(E_c/E_a)]}\left(\frac{E}{E_a}\right)^{-3/2}, 
\label{a}
\end{equation}
where $\eta$ is the effectives of the energy transformation from the $E\sim M_X/2$ range
to the $\gamma$ range of the photons energy spectra. Let us normalize the 
 energy density of injection $\dot{Q}(t)$ (${\rm eV} {\rm cm}^{-3}{\rm s}^{-1}$)  to the present
day value $\dot{Q}(t_u) $
\begin{equation}
\dot{Q}(t)=\omega(t)\dot{Q}(t_u)
\label{a1}
\end{equation}
where $t_u$ is the age of the Universe at $z=0$ and $\omega(t)\propto
\varepsilon_X$  according to
Eq.~(\ref{eqd2}). 
Integration Eq.~(\ref{a}) over all energies of the photons corresponds to the
net  of the energy density injection $\dot\epsilon_{\gamma}=\int dE E N(E)=\omega(t)\dot{Q}(t_u)$.
But for the energy range $E\ge I_{\alpha}$ the net of the ionized UV photons is given by
integral  $ \int dE N(E)$:
 \begin{equation}
\dot{n}_i\simeq \frac{\omega(t)\dot{Q}(t_u)}{I}\left(\frac{I}{E_a(z)}\right)^{\frac{1}{2}}
\label{a2}
\end{equation}
Using normalization for the energy density $\dot\epsilon_{\gamma}H^{-1}\simeq \epsilon_{\rm EGRET}$
to the EGRET energy density $ \epsilon_{\rm EGRET}\simeq 4 10^{-7} {\rm eV} {\rm cm}^{-3} $ (Sigl 2001)
 of the $\gamma$ quanta at the present moment of the cosmological expansion, we can estimate
the function $\omega(t)$. Namely, assuming 
\begin{equation}
\omega(t)\propto t^{4-p}\propto (1+z)^{\frac{3}{2}(4-p)}
\label{a3}
\end{equation}
we can find that
\begin{equation}
\varepsilon_{i}\simeq \varepsilon_{\alpha}\simeq \xi (z)
\left(\frac{ \epsilon_{\rm EGRET}}{In_b(z=0)}\right)(1+z)^{\frac{3}{2}(1-p)} 
\propto (1+z)^{2(1-\frac{3}{4}p)}
\label{a4}
\end{equation}
where $\xi(z)= (I/E_a(z))^{1/2}\simeq 3 \times 10^{-6}\sqrt{1+z}$.
The power index $p=1$ correspond the model involving release of $x$-particles
from topological defects, such as ordinary cosmic strings, necklaces and
 magnetic monopoles (see for review by Sigl 2001). The model $p=2$ corresponds to the
SHDM exponential decay (see Eq.~(\ref{eqd2})) with $\tau_x\gg t_U$. From Eq.(\ref{a4}) 
one can see that for $p=4/3$ the $\varepsilon_{i}$ and $\varepsilon_{\alpha}$ parameters
does not depend on the redshift, which is exactly the case of the model by Peebles, Seager and
Hu (2000). 

One additional possibility comes from the model with $\tau_X \le t_u$. This model could be
related with primordial black holes evaporation or all SHDM relics decay, for which the corresponding
life time is less than the present age of the Universe. Note that for $\tau_X\simeq 0.1 t_u$
(and corresponding redshift $z_X \sim 6-7$ ) the SHDM relic decay before or just at the galaxy
formation epoch and the observable flux of the UHECRs mostly should relate with
extragalactic background of the particles, which is characterized by very small anisotropies.
For such a model we need to modify Eq.~(\ref{a4}) as $\xi_{mod}(z)=\xi(z)\Theta(z)$ for
$z\gg z_X$, where
\begin{equation}
\Theta(z_X)=\exp\left[(1+z_X)^{\frac{3}{2}}\right],
\label{a5}
\end{equation}
and $z_X$ corresponds to the $\tau_X$. 

In Fig.5 I show the dependence of the $\varepsilon_{i}$ parameter on the redshift $z$ for
various models of the SHDM decay (for different $p$ and $\Theta(z_X)$).

\begin{figure}[!t]
\centering
\vspace{0.01cm}\hspace{0.1cm}\epsfxsize=10cm
\epsfbox{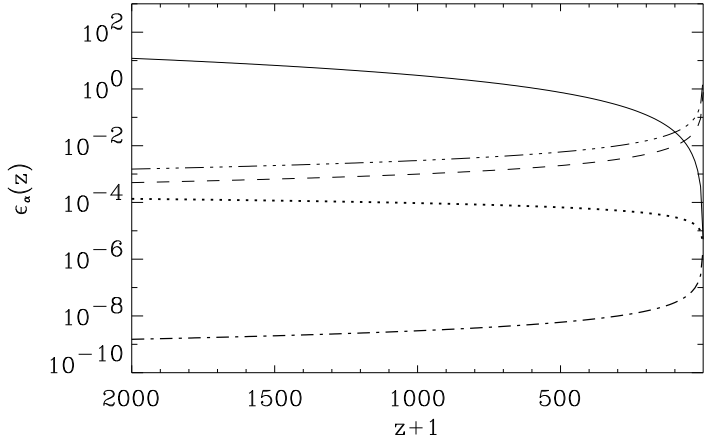}
\vspace{0.5cm}
\caption{The dependence of the $\varepsilon_{i}$ parameters versus $z$ for 
different models of the SHDM decay. Solid line corresponds to $p=0,\Theta(z_X)=1$ model,
dot line is $p=1,\Theta(z_X)=1$ model, dash dot line is the model $p=2,\Theta(z_X)=1$
and dash line and dash dabble dot line are the models
$p=2 , \Theta(z_X)=10^6$ and $p=2 , \Theta(z_X)=3 \times 10^5$  correspondingly. 
}
\end{figure}
As one can see from Fig.5, some of the SHDM relics can distort the kinetics of
hydrogen recombination quite significantly 
($\varepsilon_{\alpha},\varepsilon_i\ge 10^{-3}-10^{-2}$ for $z<1500$), for some of the
SHDM relics the perturbation of the ionization history of the Universe is small and
practically unobservable. However, all energy injection during the epoch of 
hydrogen recombination would by partially absorbed by the hydrogen atoms and
produce corresponding delay of recombination or ionization of the neutrals. 
 
The question is we can have the acceleration of the hydrogen recombination and what we
need to know about the properties of the cosmological plasma beyond the standard model? 

\subsection{Acceleration of recombination}
In the framework of non-standard recombination models, it was shown that
all peculiarities of the hydrogen photo-ionization kinetics can be described in terms of
$\varepsilon_{\alpha},\varepsilon_i$  parameters, which correspond to additional
non-equilibrium $Ly_{alpha}$ and $Ly_c$ photon injection. In the previous section
it was shown that for different models of the SHDM particles decay both parameters
$\varepsilon_{\alpha},\varepsilon_i$ are positive and hence the delay of hydrogen recombination
occurs. In terms of the $\varepsilon_{\alpha},\varepsilon_i$ parameters any acceleration
of recombination means that one or both of them are negative. However,  any decay
of the SHDM relics do not produce ``negative'' energy injection and, taking into account
the black-body CMB photons, acceleration of the
recombination means that we can remove some of the photons from interaction with the formed
neutral hydrogen atoms, which is impossible. Does it mean that for
given values of the cosmological parameters ($\Omega_b,\Omega_m$\ldots etc.), which determines
 the general properties of the model of the Universe, we can only have the delay of
hydrogen recombination? To answer this question, we need to focus 
on Eq.~(\ref{eqla6}),  in which the net of free electron number density decreases
during recombination which is  proportional to the electron and proton  number density. 
The non-linear dependence of the ionization fraction on baryonic number density
is the key to the construction of the accelerated model of recombination!

The basic idea of accelerated recombination is very transparent (Naselsky and Novikov 2002).
Suppose that, instead of pure adiabatic model of initial metric, density and velocity perturbation of     
matter we have the mix between adiabatic and isocurvature (or isotemperature) perturbation. If at the 
very small mass scales of isocurvature perturbations, e.g. those comparable with
$10^2-10^5 M_{\odot}$, the distribution of the baryons is very non-uniform 
like clouds, beyond which scale perturbations follow the standard adiabatic
properties, then the hydrogen recombination inside the clouds goes faster than outside the
clouds. This indicates that
for adiabatic perturbations which produce anisotropies and polarization of the CMB, the 
hydrogen ionization differs from the standard one, which corresponds to the mean value
of the baryonic matter density.  Non-trivial is that such a modification of the
ionization history of the hydrogen in the cloudy baryonic model can be described
in terms of $\varepsilon_{\alpha}<0$ and $\varepsilon_{i}> 0$ parameters as
for the delayed recombination model (Naselsky and Novikov 2002). 
All of these delayed and accelerated recombination models could be very important
for reconstruction of the best-fitting cosmological model parameters from the modern CMB observational    
data and can distort the optimal values.

\section{Non-standard recombination and its manifestation in the
CMB anisotropy and polarization power spectrum}
For illustration of the introduced above classification of the  hydrogen recombination regimes we will use
two cosmological models with the following parameters. The Model 1: $\Omega_bh^2=0.022,\Omega_c=0.125, 
\Omega_K=0 ,\Omega_{\Lambda}=0.7, \tau_r=0.1, h=0.7, n_s=1,\varepsilon_{\alpha}=\varepsilon_{i}=0$.
This model has the best-fitting of the CBI observational data (Mason et al 2002). 
The Model 2: $\Omega_bh^2=0.032,\Omega_c=0.115, 
\Omega_K=0 ,\Omega_{\Lambda}=0.7, \tau_r=0.1, h=0.7, n_s=1,\varepsilon_{\alpha}=\varepsilon_{i}=0$.
This model differs from the previous one on the baryonic and CDM fraction densities and it is non-optimal
according to the CBI data. Using RECFAST code we can find the differences in the ionization fraction of hydrogen
plotted in Fig. 6.
\begin{figure}
\centering
\vspace{0.01cm}\hspace{-0.1cm}\epsfxsize=10cm
\epsfbox{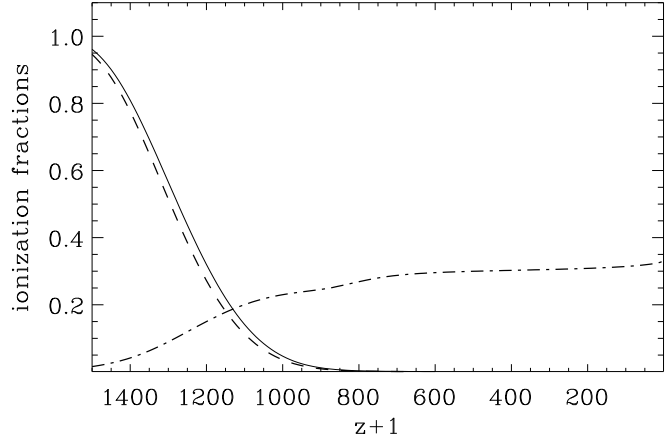}
\vspace{0.5cm}
\caption{The rate of ionization  for the  Model 1 and 2. Solid line corresponds to the Model 1 ($x_1$),
dash  line Model 2 ($x_2$), dash-dot line the ratio $(x_1-x_2)/x_1$ . The reionization epoch are not
included. }
\end{figure}

Fig.7 is the plot of the ionization fraction of the hydrogen in the clumpy baryonic model (Model 3), which corresponds
to the parameters of the Model 1, but instead of $\Omega_bh^2=0.022$ we have to use the mean value of the
baryon density $\langle \Omega_b \rangle h^2=0.022$, assuming that density contrast between inner and outer regions
$\xi=\rho_{in}/\rho_{out}=11$, the fraction of volume in clouds $f=0.1$ and the mass fraction
$Z=\xi f/[1+(\xi-1)f]=0.5$.
\begin{figure}
\centering
\vspace{0.01cm}\hspace{-0.1cm}\epsfxsize=10cm
\epsfbox{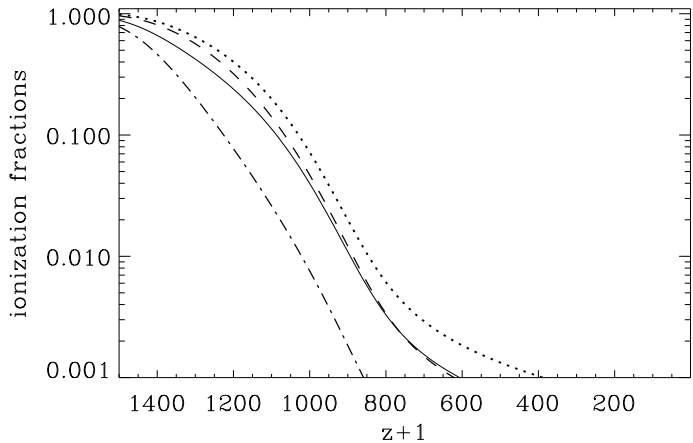}
\vspace{0.5cm}
\caption{The rate of ionization for the clumpy baryonic  model. 
 Dash-dot line corresponds to the ionization
fraction for the inner regions at. Dot line is the ionization fraction for the outer regions. Solid line the mean
fraction of the ionization. Dash line is the ionization fraction for the model with
$\Omega_{b}=0.045$. }
\end{figure}

The mass spectrum of a clouds means to be close to $\delta(M-M_{cl})$, where
$10^2 M_{\odot}<M_{cl}< 10^6 M_{\odot}$ . For this model the mean ionization fraction is related with
ionization fractions inside $x_{in}$ and outside $x_{out}$ clouds as $x_{\rm mean}= x_{in}Z +x_{out}(1-Z)$
(Naselsky and Novikov 2002). In Fig.8 we show the difference between  $x_{\rm mean}$ and $x_1$ normalized to
$(x_1 + x_{\rm mean})/2$.
\begin{figure}
\centering
\vspace{0.01cm}\hspace{-0.1cm}\epsfxsize=10cm
\epsfbox{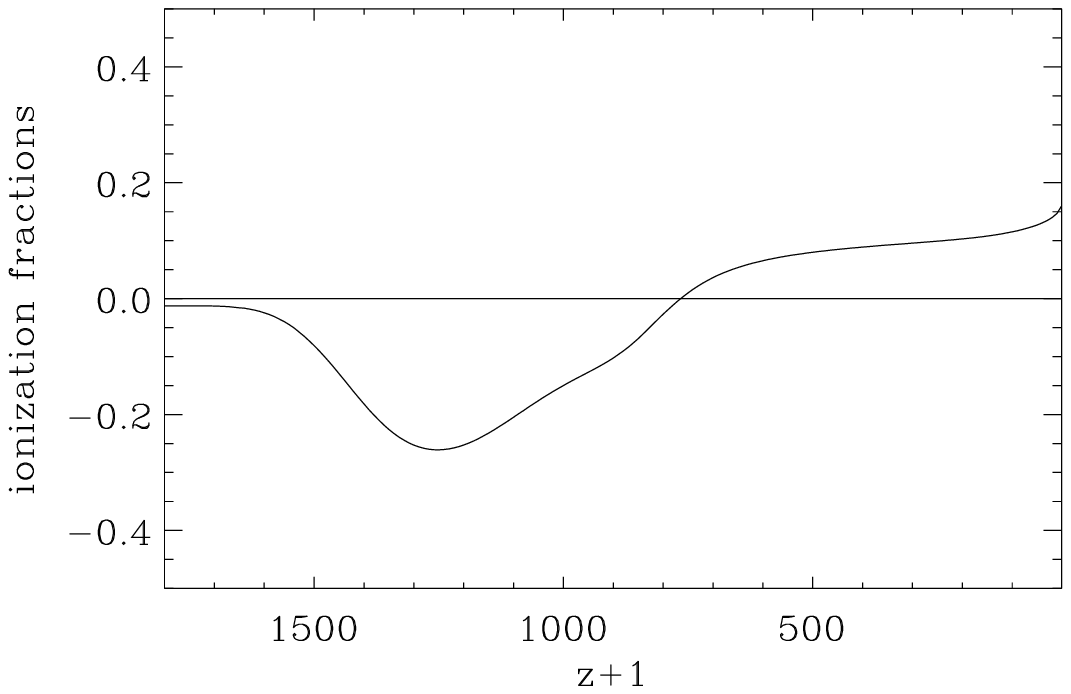}
\vspace{0.5cm}
\caption{The difference of the rates of ionization between the clumpy baryonic cosmological model 
and without clumps.}
\end{figure}

As one can see from Fig.8, at the range $700 < z < 1500$ we have acceleration of recombination while
at the range $z < 700$ we have delay of recombination. 
The last model, Model 4,  which we would like to include in our analysis is the Model 2 plus additional 
sources of delay and ionization, corresponding to the following values of the 
$\varepsilon_{\alpha} , \varepsilon_{i}$ parameters:
\begin{eqnarray}
\varepsilon_{\alpha}(x)&=&\alpha\times exp\left[-\left(\frac{1+z}{1+z_x}\right)^{-\frac{3}{2}}\right] 
\left(\frac{1+z}{1+z_x}\right)^{-\frac{3}{2}} ; \nonumber \\
\varepsilon_{i}(z)&=&\frac{\beta}{\alpha} \varepsilon_{\alpha}(z)
\label{eq6}
\end{eqnarray}
where $\alpha\simeq 0.3$, $\beta\simeq 0.04$ and $z_x\simeq 10^3$. As one can see from Eq(6) at 
$z\simeq10^3 $ the variation of the $\varepsilon_{\alpha} , \varepsilon_{i}$-parameters are very slow
and $\varepsilon_{\alpha}\simeq \alpha$, $\varepsilon_{i}\simeq\beta$for comparison with the parameters by
 Peebles, Seager and Hu (2000). In Fig.9 I plot the ionization fraction in Model 4 in comparison with
ionization fraction in the Model 1 and the Model 2. As one can see the delay of recombination in the 
Model 4 produce $20-30 \%$ deviation of the ionization fraction at the redshift $z\simeq 10^3$ and
very significant deviations at the redshift $z< 10^3$. 

\begin{figure}
\centering
\vspace{0.01cm}\hspace{-0.1cm}\epsfxsize=10cm
\epsfbox{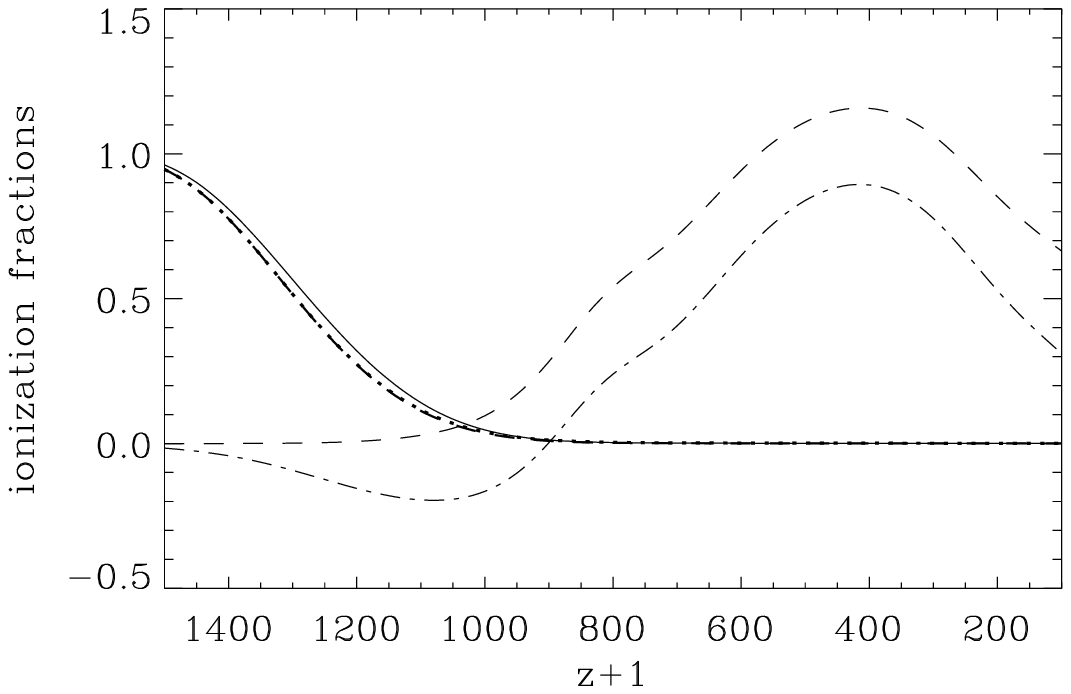}
\vspace{0.5cm}
\caption{The rates of ionization and the differences of the rates of ionization between the 
Model 1, 2 and 4. Solid line is the ionization fraction in the  Model 1. Dotted and short-dash line are 
the Model 2 and 4, respectively, which look very similar. Long-dash line corresponds to the ratio
$2(x_4-x_1)/ (x_1+x_4)$. Dash-dot line corresponds to $2(x_4-x_2)/ (x_2+x_4)$.}
\end{figure}

It is worth noting that all the Models 1-4 clearly illustrate the different delayed and accelerated regimes,
discussed at the beginning of the this section. The question is how sensitive the CMB anisotropy and polarization
power spectrum on different models of ionization history of hydrogen and baryonic density?
Below I will show that combination $\Omega_b,\varepsilon_{\alpha},\varepsilon_{i}$ is very important
for estimation of the correct value of the baryonic fraction density from the CMB data.

\subsection{Anisotropy and polarization as a test on ionization regimes}

In order to compare the CMB anisotropy and polarization power spectrum in the Models 1-4 we have to use
modification of the CMBFAST code (Seljak and Zaldarriaga 1996), by taking into account more complicated   
ionization history of the plasma.
All the models which include the late reionization correspond to the standard CMBFAST option 
for a given value of the optical depth $\tau_r$. This value of $\tau_r$ is 
related with the ionization fraction $x=1$ at the redshift of reionization $z_r$ as
\begin{equation}
 z_{re}=13.6\left(\frac{\tau_r}{0.1}\right)^{2/3}\left(\frac{1-Y_p}{0.76}\right)^{-2/3} 
\times\left(\frac{\Omega_b h^2}{0.022}\right)^{-2/3}\left(\frac{\Omega_c h^2}{0.125}\right)^{1/3}
\label{eq7}
\end{equation}
where $Y_p$ is the helium mass fraction of the matter. In Fig.10 we plot the CMB power spectrum for
the Models 1-4 in comparison with the BOOMERANG, MAXIMA-1, and CBI data at the multipole range
$\ell \le 2000$ in which the possible manifestation of the Sunuyaev-Zel'dovich effect is not important. 
\begin{figure}
\centering
\vspace{0.01cm}\hspace{-0.1cm}\epsfxsize=8 cm
\epsfbox{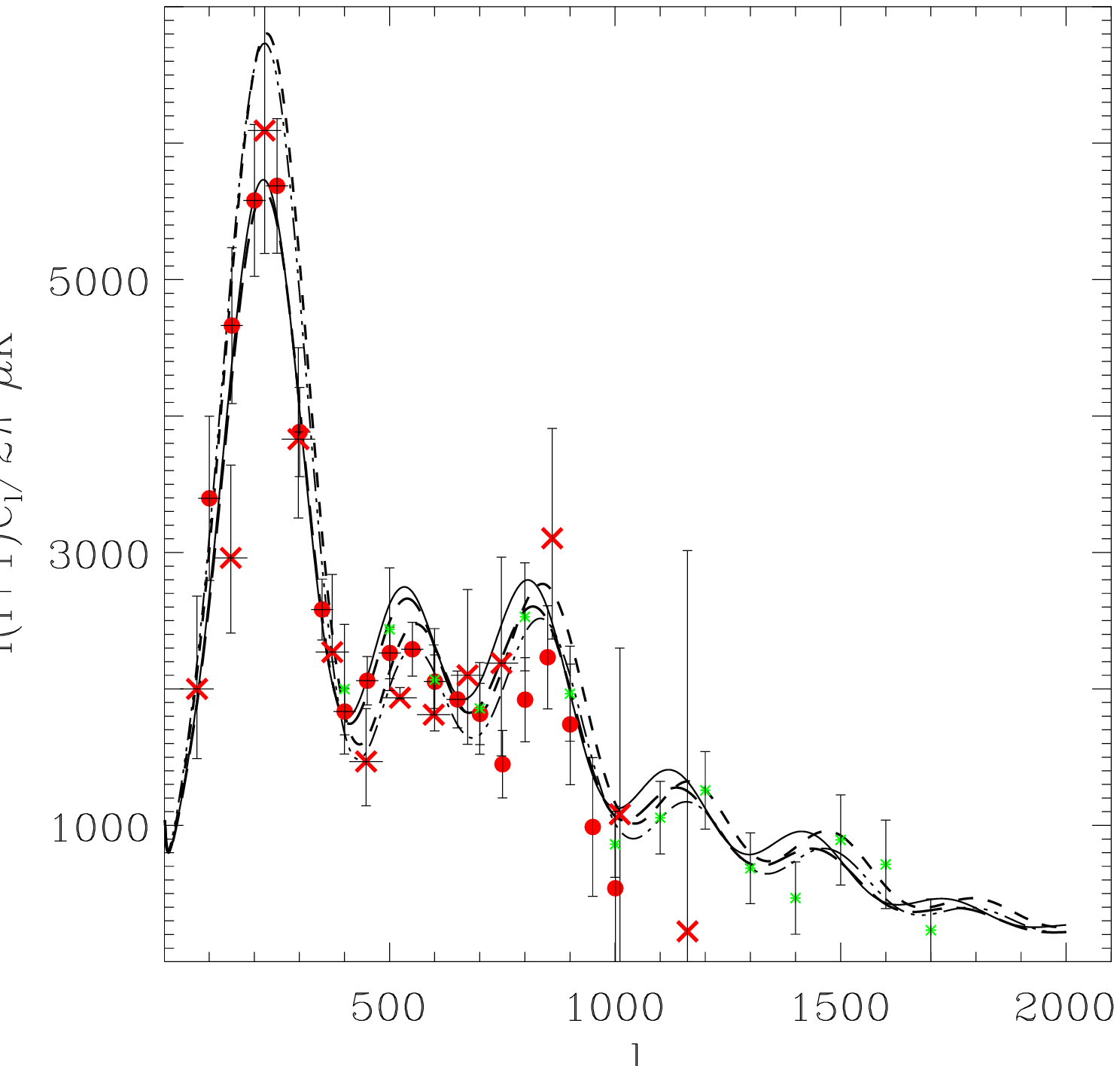}
\vspace{0.5cm}
\caption{The CMB power spectrum for the Models 1-4. Solid line corresponds to the Model 1,
 dash line the Model 2, dash-dot line the Model 3, and long-dash line the Model 4.}
\end{figure}

\begin{table}[]
\caption{$\chi^2$ for observations and models}
\label{tbl2}
\begin{tabular}{lcc ccc } 
  obs/mod &BOOM&MAXIMA& CBIM1 & CBIM2 & VSA\cr
\hline
  Model 1 &11.2 &14.1 & 3.20 & 7.15 & 7.08\cr
  Model 2 &34.6 &17.7 & 2.75 & 7.51 &13.0\cr
  Model 3 &8.69 &11.3 & 1.60 & 5.62 & 6.59\cr
  Model 4 &27.7 &15.7 & 1.87 & 5.15 & 10.0\cr
\hline
\end{tabular}
\end{table}

As one can see from Fig.10 all the models look very similar to each other. However, in the Table 1
we show the values of the $\chi^2$-parameter for  all  the models. From the Table 1 we seen that
the Model 3  with baryonic clouds has an excellent agreement with the CBI data while the Model 1
is characterized by the value of the $\chi^2_1=2\chi^2_3$. Note that the Model 1 is the best-fitting      
 model for the CBI data without any assumption on more complicated ionization history of
cosmological hydrogen recombination. The Model 4 corresponds to the $\Omega_b h^2$ parameter
1.5 times bigger than the Model 1. However, this model also has an excellent agreement with all
the CBI observational data.

From the practical point of view, the Models 1, 3 and 4
consistent with the CBI data and more complicated ionization history of the hydrogen recombination
manifest themselves as new sources of degeneracy of the cosmological parameters, namely, the
baryonic density of matter. This conclusion is very important for the upcoming
MAP and \planck data.
In order to characterize the differences
between the models and to compare their with the sensitivity of the upcoming \planck experiment 
we plot in Fig.11 the functions
 \begin{eqnarray}
 D_{1,3}(\ell)=2\left[C_1(\ell)-C_3(\ell)\right]/\left[C_1(\ell)+C_3(\ell)\right]; \nonumber \\
 D_{2,4}(\ell)=2\left[C_2(\ell)-C_2(\ell)\right]/\left[C_2(\ell)+C_2(\ell)\right];
\label{eq8}
\end{eqnarray}
for the multipole range $2< \ell < 2500$. Obviously, these functions $ D_{1,3}(\ell)$ and  $ D_{2,4}(\ell)$
is necessary to compare with the errors of the
$\Cl$ extraction for the \planck mission. 

\begin{figure}
\centering
\vspace{0.01cm}\hspace{-0.1cm}\epsfxsize=10cm
\epsfbox{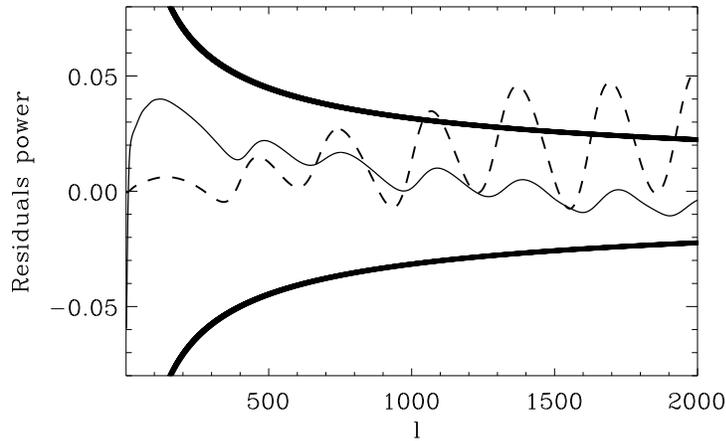}
\vspace{0.5cm}
\caption{The differences in the CMB power spectrum for the Models 1-4. Solid line corresponds to the 
$D_{1,3}(\ell)$ function and dash line  $D_{2,4}(\ell)$. Thick solid line is the estimated errors for the 
\planck. }
\end{figure}

\begin{figure}
\centering
\vspace{0.01cm}\hspace{-0.1cm}\epsfxsize=10cm
\epsfbox{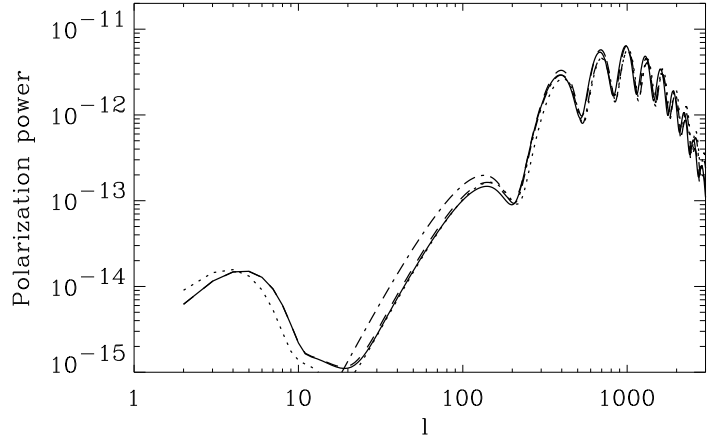}
\vspace{0.5cm}
\caption{The CMB polarization power spectrum for the Models 1-4. Solid line corresponds to the Model 1,
 dot line the Model 2, dash line the Model 3, and  dash-dot line the Model 4. }
\end{figure}

\begin{figure}
\centering
\vspace{0.01cm}\hspace{-0.1cm}\epsfxsize=10cm
\epsfbox{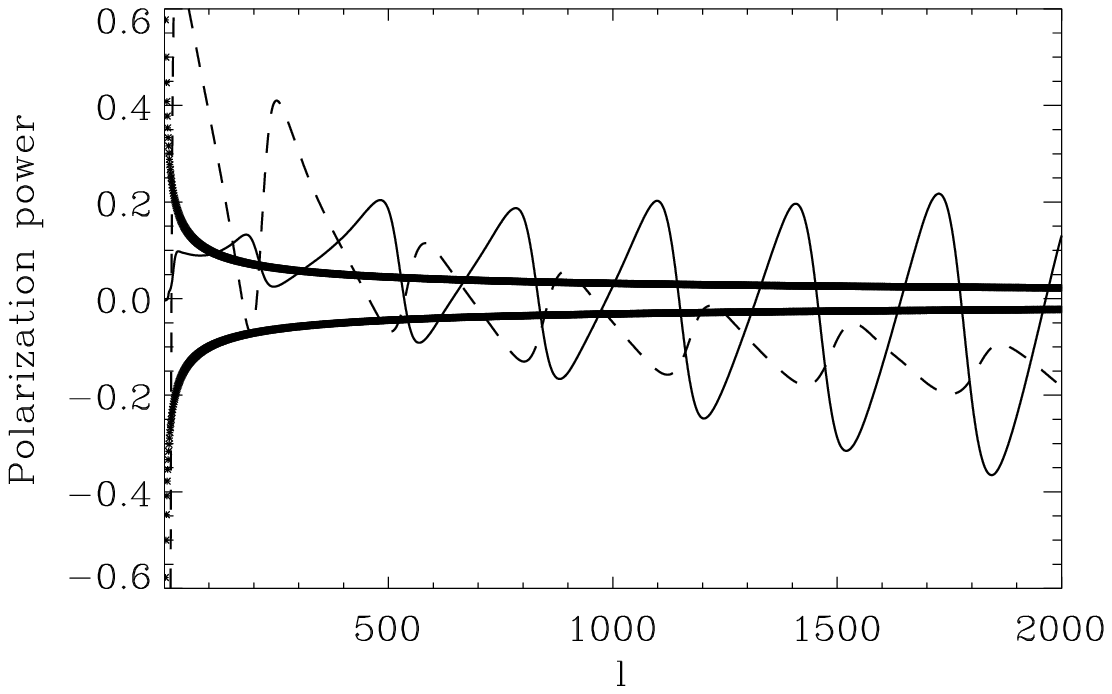}
\vspace{0.5cm}
\caption{The differences of the CMB polarization power spectrum for the Models 1-3 and the Models 2-4. Definition for 
residuals are similar to the definition for the anisotropy. Solid line is  for $D_{1,3}(\ell)$ and dash
line $D_{2,4}(\ell)$, but for polarization. Thick solid line is the estimated errors for the \planck. }
\end{figure}

As one can see from Fig. 11,  the range of the  $ D_{1,3}(\ell)$ and  $ D_{2,4}(\ell)$ values is close to
the $5\%$ . Note that it can  easily drop down to $1-2\%$ using smaller values of the
parameters of the non-standard ionization history   $\varepsilon_{\alpha} , \varepsilon_{i}$ and
corresponding parameters $\xi$, $Z$, $z_X$, $m$ and so on. These parameters play a role of the
``missing parameters'' in the model of the CMB anisotropy formation and  significantly increase
the number of the standard parameters $\Omega_b$, $\Omega_{\Lambda}$,\ldots $\Omega_K$.
It is well known that for analysis of the degeneracy of some of the cosmological parameters the
polarization measurements are very useful. In Fig.7 we plot the polarization power spectrum
$\Delta T^{2}_ p (\ell)=\ell(\ell+1)/2\pi C_p(\ell)$ for the Models 1-4. As one can see from this Figure, the difference
of the $\Delta T^{2}_ p (\ell)$ is very small at the range $\ell>200$ and quite significant for the Model 1 and 2 relatively
the Models 3 and 4 because of the bump at $\ell \sim 5-10$, which is related with reionization
of the hydrogen at the epoch of formation of galaxies and clusters of galaxies. Some useful
information can be obtained at the range $20 <\ell <200$ in which the manifestation of the Model 4
 is significant. This Model shows that the optical depth $\tau_r=0.1$ can be obtained without
distortions of the low multipole part of the power spectrum at $\ell\sim 5-10$. However, for the \planck
mission one of the most important source of uncertainties at the multipole range $\ell \le 200$ is the
the cosmic variance. The corresponding errors of the $\Delta T^2_p(\ell)$ estimation at this range is
$\delta C_p(\ell)/C_p(\ell)\simeq (f_{\rm sky}\ell)^{-1/2}\simeq 9\left(\ell/200\right)^{-1/2}
\left(f_{\rm sky}/0.65\right)^{-1/2}{\%}$.
\footnote{We assume that instrumental noise and systematics should be close
to the cosmic variance limit .} 
For $f_{\rm sky}/0.65\simeq 1$ and $\ell \simeq 2000$ the cosmic variance limit on
$\delta C(\ell)/C(\ell)$ is $3\%$ at $68\%$ confidence level, which means
that all the peculiarities of the anisotropy power spectra below this
limit should be unobservable. For the CMB polarization, as it follows from 
Fig.8 all features of the power spectrum can be observed by the \planck and probably, the low multipole   
part of the power spectrum can be observed by the MAP mission.
So future experiments for detection of polarization can be very useful for investigation
of the ionization history of the cosmic plasma and possible distortions of the kinetics of 
hydrogen recombination by the different sources.

\section{Conclusion}
The precision of the measurements on the CMB anisotropy and polarization, especially being expected from  
the MAP and \planck missions, allows us to distinguish the cosmological  
models under consideration and to discuss $10-20\%$ distortions of the
standard recombination process occurring at redshift$z\simeq 10^3$. 

In my talk I have compared available data from observations of the CMB anisotropies with 
several cosmological models with different kinetics of hydrogen recombination, which is caused
 by possible external sources of the energy injection and possible small-scale clustering of the baryonic
component in the clouds. It was shown that such a sources can produce the delay or
acceleration of hydrogen recombination and increases the number of parameters
used to fit observed power spectra of the CMB anisotropy and polarization.

The most interesting conclusion coming from the investigation of the non-standard
models of hydrogen recombination is that the CMB polarization measurements 
will play an important role for the estimation of the ``missing parameters''
from the upcoming \planck data. 

\section*{Acknowledgment}
I would like to thank Prof. Norma Sanchez for the invitation to participate this School. 
This paper was supported in part by Danmarks Grundforskningsfond 
through its support for the establishment of the Theoretical 
Astrophysics Center, and by grants RFBR 17625.

I am grateful to L.-Y. Chiang, A. Doroshkevich,  
I. Novikov and A. A. Starobinsky 
for discussions and help during the preparation of this article.

\section{References}
Abu-Zayyad et al., 2001, ApJ, 557,  686 \\
Adams, J.A., Sarkar, S. \& Sciama D.W., 1998,  MNRAS, 301, 210\\
Ave, M.,et al., Phys. Rev. Lett.,  85, 2244, 2000;\\ 
Avelino, P., Martins, C., Rocha, G., \& Viana, P., 2000, Phys. Rev. D, 62, 123508\\
Battye, R., Crittenden, R., \& Weller, J., 2001, Phys. Rev. D, 63, 043505\\
Bhattacharjee P. \& Sigl, G. 2000, Phys. Rept.,  327, 109\\
Berezinsky, V. S., Bulanov, S. V., Dogel, V. A., Ginzburg, V. L. and V. S. Ptuskin,
Astrophysics of Cosmic Rays, (North Holland, Amsterdam,1990)\\
Berezinsky, V. S., Kashelrie$\ss$,  M. \& A. Vilenkin, 1997, Phys. Rev. Lett, 
 79, 4302, \\
Berezinsky, V.S., Blasi, P. and  A. Vilenkin, 1998, Phys. Rev. D,  58, 
103515\\
Birkel M. and  Sarkar, S. 1998, Astropart. Phys.,  9, 297\\
de Bernardis, P., et al., 2000, Nature, 404, 955\\
Blasi,P.,1999,  Phys.Rev.D,  60, 023514, .\\ 
Doroshkevich A. G., \& Naselsky, P. D., 2002, Phys. Rev. D, 65, 123517 \\
Doroshkevich A. G., Naselsky, I.P.., Naselsky, P. D. \& Novikov, I.,D., 2002, astro-ph/0208114 \\
Ellis, J., Gelmini, G., Lopez, J., Nanopoulos, D., \& Sarkar, S., 1992, Nucl. Phys. B, 373 , 399\\
Halverson, N. W., 2002, ApJ, 568, 38 \\
Hayashida, N., et al., 1994, Phys. Rev. Lett,  73, 3491.\\
Hanany, S., et al., 2000, ApJ, 545, L5\\
Greisen, K. 1966, Phys. Rev. Lett.,  16, 748.\\
Jones, B. J. T., \& Wyse R., 1985, A\&A, 149, 144\\
Yoshida, S. \& H. Dai, 1998., J. Phys.G,  24, 905.\\
Ivanov,P.B., Naselsky, P.D. and Novikov, I.D., 1994,   Phys. Rev.D,50,7173.\\
Kotok , E.V. and Naselsky, P.D.,1998, Phys. Rev.D, 58,103517.\\
Kovac, J., et al., 2002, astro-ph 0209478.\\
Kuzmin, V.A., \& V.A. Rubakov, 1998, Yader.Fiz.,  61, 1122\\
Landau, S., Harari, D., \& Zaldarriaga, M., 2001, Phys. Rev.D, 63, 3505 \\ 
Lawrence, M. A., Reid, R. J. O. \& Watson, A. A., 1991, J. Phys. G., 
Nucl.Part.Phys.,  17, 733\\
Mason, B. S., et al., 2002, astro-ph/0205384\\
Naselsky, P. D., 1978, Sov. Astron. Lett., 344, 4\\
Naselsky, P. D., \& Polnarev, a. G., 1987,  Sov. Astron. Lett., 13, 67 \\
Naselsky, P.D.,\& Novikov, I.,D.,2002, MNRAS, 334, 137 \\
Peebles, P., 1968, ApJ, 153, 1\\
Peebles, P., Seager, S., \& Hu, W., 2000, ApJ, 539, L1 \\
Protheroe, R.J., \& Biermann, P. L., 1996, Astrop. Physics,  6, 45\\
Protheroe, R.J., Stanev, T. , and V.S.Berezinsky, 1995, astro-ph/9409004
\\
Rybicki, G. B and Lightman, A. P., 1979, Radiative processes in astrophysics,
A Wiley-Interscience publication\\
Sarkar, S., and  Toldra, R., 2002, Nucl. Phys., B621, 495\\
Sarkar, S., \& Cooper, a., 1983,  Phys. Lett. B, 148, 347\\
Scott, D., Rees, M.,J., \& Sciama, D., W., 1991, A\&A, 250, 295\\
Seager , S., Sasselov, D., \& Scott, D., 1999, ApJ, 523, L1\\
Seljak, U., \& Zaldarriaga, M., 1966, ApJ, 469, 437\\
Sigl, G., 2001, hep-ph/0109202\\
Tegmark, M., 2001, A\&AS, 199, 3407\\
Zabotin, N. A., \& Naselsky, P. D., 1982, Sov. Astron., 26, 272\\ 
Zatsepin G.T., and Kuzmin, V.A.,1966, Pis'ma Zh. Eksp.Theor.Fiz., 
 4, 114,; JETP. Lett., 4, 78, 1966.\\
Zel'dovich, Ya. B., Kurt, V.,\& Sunuyaev, R.A., 1968, Zh. Eksp. Theor. Phys., 55, 278\\
Watson, R. A., et al., 2002, astro-ph/0205378
\end{document}